\documentclass[twocolumn]{aastex61}

\usepackage{amsmath,amssymb,amsthm,mathrsfs,dsfont}
\usepackage{multirow}
\usepackage{graphicx}
\usepackage{booktabs}
\usepackage{epstopdf}
\usepackage{color}
\usepackage{hyperref}

\usepackage{stmaryrd}
\usepackage{ulem}

\def\be{\begin{equation}}
\def\ee{\end{equation}}
\def\bd{\begin{displaymath}}
\def\ed{\end{displaymath}}
\def\ba{\begin{aligned}}
\def\ea{\end{aligned}}
\def\nms{\mathsurround=0pt}
\def\oversim#1#2{\lower 4pt\vbox{\baselineskip 0pt \lineskip 1pt
    \ialign{$\nms#1\hfil##\hfil$\crcr#2\crcr\sim\crcr}}}
\def\arcdeg{{^{\circ}}}
\def\tot{{\rm tot}}
\def\rg{r_{\rm g}}
\def\cm{{\rm cm}}

\def\LOS{{\rm LOS}}

\def\bh{M_{\bullet}}

\def\g{{\rm g}}

\def\msun{M_{\odot}}

\def\max{{\rm max}}
\def\min{{\rm min}}

\begin{document}

\title{X-ray Eclipses of Active Galactic Nuclei} 
\shortauthors{Zhang, Yu, \& Lu}
\author{Fupeng Zhang$^\ast$}
\affiliation{Kavli Institute for Astronomy and Astrophysics, Peking
University, Beijing, 100871, China; $^\dagger$\,yuqj@pku.edu.cn}
\affiliation{School of Physics and Astronomy, Sun Yat-Sen University,
Guangzhou 510275, China; $^\ast$\,zhangfp7@mail.sysu.edu.cn;}
\author{Qingjuan Yu$^\dagger$}
\affiliation{Kavli Institute for Astronomy and Astrophysics, Peking
University, Beijing, 100871, China; $^\dagger$\,yuqj@pku.edu.cn}
\author{Youjun Lu$^\ddagger$}
\affiliation{National Astronomical Observatories, Chinese Academy of
Sciences, Beijing, 100012, China; $^\ddagger$\,luyj@nao.cas.cn}
\affiliation{School of Astronomy and Space Sciences, University of 
Chinese Academy of Sciences, No. 19A Yuquan Road, Beijing 100049, China}

\begin{abstract}
X-ray variation is a ubiquitous feature of active galactic nuclei
(AGNs), however, its origin is not well understood.  In this paper, we
show that the X-ray flux variations in some AGNs, and
correspondingly the power spectral densities (PSDs) of the variations,
may be interpreted as being caused by absorptions of eclipsing clouds or
clumps in the broad line region (BLR) and the dusty torus.  By
performing Monte-Carlo simulations for a number of plausible cloud
models, we systematically investigate the statistics of the X-ray
variations resulting from the cloud eclipsing and the PSDs of the
variations. For these models, we show that the number of eclipsing
events can be significant and the absorption column densities due to
those eclipsing clouds can be in the range from $10^{21}$ to $10^{24}
{\text{cm}}^{-2}$, leading to significant X-ray variations. We find
that the PSDs obtained from the mock observations for the X-ray flux
and the absorption column density resulting from these models can be
described by a broken double power law, similar to those directly
measured from observations of some AGNs. The shape of the PSDs depend
strongly on the kinematic structures and the intrinsic properties of
the clouds in AGNs. We demonstrate that the X-ray eclipsing model can
naturally lead to a strong correlation between the break
frequencies (and correspondingly the break timescales) of the PSDs and
the masses of the massive black holes (MBHs) in the model AGNs, which
can be well consistent with the one obtained from observations. Future
studies of the PSDs of the AGN X-ray (and possibly also the optical-UV) flux and
column density variations may provide a powerful tool to constrain the structure
of the BLR and the torus and to estimate the MBH masses in AGNs.
\end{abstract}


\section{Introduction} 
X-ray variation is a ubiquitous feature of active galactic nuclei
(AGNs; e.g., \citealt{Mushotzky93}). The variation timescale of the
AGN X-ray emission ranges from hours, days, to years
\citep[e.g.,][]{Mchardy87, Mushotzky93, Markowitz03}. Observations have
revealed that the power spectral density (PSD) of the X-ray variation as a
function of the variation frequency $\nu$ in many AGNs, $P(\nu)$,
can be described by a double power law with a break frequency
of $\nu_{\rm B}$, i.e., $P(\nu)\propto
\nu^{\gamma_h}$ at high frequencies ($\nu\gg\nu_{\rm B}$) with
$\gamma_h \la -2$, and $P(\nu)\propto \nu^{\gamma_l}$ at low frequencies ($\nu
\ll\nu_{\rm B}$) with $\gamma_l \sim -1$ \citep[e.g.][]{Mchardy87,
Uttley02, Markowitz03, GonzlezMart12}. It has been demonstrated that
the characteristic timescales corresponding to the break frequencies
($\sim 1/\nu_{\rm B}$) strongly correlate with the masses of the central
massive black
holes (MBHs) in those AGNs \citep{McHardy06, GonzlezMart12}. The properties
of some X-ray
binaries (XRBs) could also fit into this correlation
\citep[e.g.,][]{McHardy06}, but the reliability of such a
relationship is not clear, yet \citep[e.g.,][]{Done05,
Kording07}. The physics behind the break frequency versus BH mass correlation
is not well understood so far.

A number of models have been proposed to explain the observed X-ray
variation of AGNs. In general, these models can be divided into the two
main categories: (1) the variation is due to changes in the intrinsic
X-ray emission; and (2) the variation is due to changes in the materials or
clouds along the line of sight (LOS) to the X-ray emitting source that absorb
part of
the intrinsic X-ray emission. In the first category, the X-ray
variation could be due to the change of the accretion rate, and
consequently, to the change of the total number of seed photons to be
inverse Compton-scattered up to X-ray photons by the hot corona located in
the vicinity of the central MBHs of AGNs
\citep[e.g.,][]{Lyubarskii97, Lamer03a, Uttley02, Zdziarski03}, or to the
change of the hot corona itself (including its motion and location),
or to the inflation of the magnetic flares being injected into the corona
\citep[e.g.,][]{PF99, LY01a, Fabian03, Marinucci14}. In the second
category, the X-ray variation is dominated by the absorption of clouds
in multiple zones that (partially) cover the X-ray emitting source on
the observer's sky plane \citep[e.g.,][]{Lamer03b, Turner09, Miller08,
Miller09, Parker15,Abrassart00}.  
In the standard AGN unification model \citep{Antonucci93}, the natural
sources for the absorption are some line emission clouds in the
broad line region (BLR) or some clumps in the dusty torus that happen
to cross the LOS; hereafter, we refer to all of them as absorption
clouds for simplicity, unless otherwise stated.
This kind of absorption model
appears to be able to explain well not only the X-ray variations of
some AGNs, such as MCG-6-30-15, NGC 4395, NGC 4151, and NGC 1365
\citep{Turner09, Parker15}, but also the relative lack of
variability at the higher energy band of the X-ray emission
\citep[e.g.,][]{Miller08}. The X-ray variation patterns and
consequently the PSDs of the variations resulted from these two model
categories should be distinguishable from each other since their
physical origins are quite different.
The study of the origin of X-ray variations would provide an insight into
the structure of and the radiation mechanisms in the central engine of AGNs.

In the scenario where the observed X-ray variations for some AGNs are mainly
caused by absorption, when a cloud crosses the LOS, a fraction of the X-ray
photons are absorbed
depending on the physical properties of the cloud, and the rest penetrate
through the cloud and are received by the distant observer. For such an event,
we denote it as an ``X-ray eclipse'' in this paper.  The duration of an X-ray
eclipse event and the period of the events are affected by the velocity of the
cloud and its distance to the central engine.
For example, the clouds located at large distances may lead to long-term
variations because of their relatively low velocities, while those located at
small distances may lead to short-term variations.
If many clouds spreading over a large range of
distances can lead to X-ray eclipses, the X-ray variation timescales or
frequencies will
depend on both the geometric and the kinematic distributions of those clouds 
and the physical properties of individual clouds.
The PSD of the X-ray variation curves describes the distribution of power
into the variation frequency components composing the X-ray variations, and
the analysis of the PSD provides a powerful tool to statistically study the
kinematical and physical conditions of those eclipsing clouds located at
different spatial regions and further their parent populations
(clouds in the BLR and clumps in the dusty torus of AGNs).

In this paper, we construct an X-ray cloud eclipsing model, and this
model generates many observational properties of X-ray variations in AGNs,
including the PSD and the break frequency versus BH mass correlation mentioned
above.
We use Monte-Carlo
simulations to consider the kinematical motion of the clouds and
clumps in the BLR and the dusty torus, and realize the X-ray eclipsing
events over a long period and consequently generate mock observations
of the X-ray flux variation. We also investigate the dependence of the
PSDs obtained from the mock observations on the kinematic structure
and the intrinsic properties of the clouds. 
Note here that we only consider those AGNs in which the X-ray
variations are dominated by the absorption of eclipsing clouds. For simplicity,
we do not intend to simultaneously consider the X-ray variations caused by the
changes in the intrinsic X-ray emission, which may
dominate the detected variations in some other AGNs.

The paper is organized as follows.  In Section~\ref{sec:analytic}, we
construct the X-ray eclipsing model and analyze
the event rate and the properties of the X-ray eclipses by
assuming that the eclipsing clouds are from the BLR and the dusty
torus. We investigate the dependence of the rate and the properties of
the X-ray eclipses on the kinematic and spatial structure of the
parent population of the eclipsing clouds and the intrinsic properties
of those clouds. By adopting a number of different models for the
spatial distribution and the properties of the clouds, we perform some
Monte-Carlo simulations to realize X-ray eclipses and generate mock
X-ray light curves in Section~\ref{sec:num_simu}.  According to those
mock X-ray light curves, we obtain their PSDs in Section~\ref{sec:pds}
and we find that they are compatible with the reasonable parameter ranges
for the spatial and kinematical distributions and
the physical properties of the eclipsing clouds. We also demonstrate
that a strong correlation between the break frequency of the PSD and
the MBH mass is a natural result of the scenario where the X-ray
variation is dominated by the absorption of eclipsing clouds, if the
inner boundary for the spatial distribution of the absorption clouds
and some intrinsic properties of those clouds scale (linearly) with the
MBH mass. Discussions and conclusions are given in
Section~\ref{sec:discussion}.

In this paper, given a set of physical variables $\mathbf{X}=(X_1,...,X_k)$
(e.g., semimajor axis, eccentricity), the probability distribution function
(PDF) of $\mathbf{X}$
is denoted by $f_{\mathbf{X}}(\mathbf{X})$ so that
$f_{\mathbf{X}}(\mathbf{X})d\mathbf{X}$ represents the number fraction of
clouds with the variable $X_i$ being in the range $X_i\rightarrow X_i + dX_i$
($i=1,..,k$) with $\int f_{\mathbf{X}}(\mathbf{X})d\mathbf{X}=1$, where
$d\mathbf{X}\equiv dX_1...dX_k$.

\section{Model for the eclipsing of X-ray emission}
\label{sec:analytic}

X-ray emission from AGNs received by a distant observer may vary due
to absorptions by clouds crossing the LOS, as revealed by observations
of some type 1 and type 2 Seyfert galaxies. The variation timescales of
the X-ray emission and the possible locations of the absorption clouds
cover a wide range as follows.
(1) Some of those absorption
events have durations of about a few hours to a few days and absorption
column hydrogen densities $\sim 10^{22} - 10^{24}\cm^{-2}$, which are probably
due to some clouds in the BLR with number densities of
$10^9-10^{11}\cm^{-3}$ and distance $\sim 10^3-10^4\rg$ from the
central MBH, where $\rg\equiv G\bh/c^2$ is the gravitational radius of
the MBH with mass $\bh$ (e.g., \citealt{Lamer03b, Maiolino10, Sanfrutos13,
Bianchi09, Markowitz14, Risaliti05a, Risaliti07, Risaliti09,
Puccetti07}). (2) Some other events have durations up to several months,
substantially longer than those due to the BLR clouds, and absorption column
hydrogen densities $\sim 10^{22} - 10^{23}\cm^{-2}$, which may be attributed to
the clumps located in the outer dusty torus with number densities of
$\sim 10^7 - 10^8\cm^{-3}$ \citep[e.g.][]{Rivers11, Marinucci13,
Miniutti14, Agsgonzlez14, Markowitz14}. Based on those
observations, we introduce simple  models below to systematically
study the eclipsing of X-ray emission from AGNs.

\subsection{Spatial Distribution of Clouds}
\begin{figure}
\centering
\includegraphics[scale=0.6]{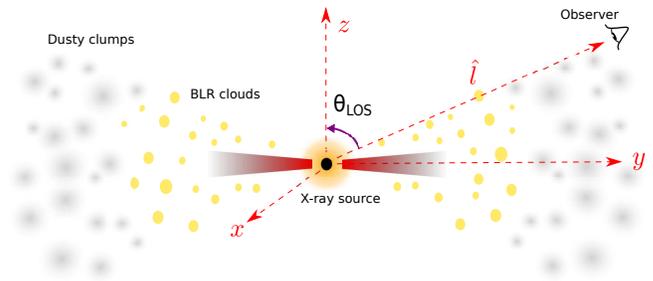}
\caption{
An illustration diagram for the spatial distribution of clouds (as a
population) surrounding the central engine of an AGN. 
The black dot at the center represents
the central MBH, and the two thick shaded bars along the y-axis
represent a thin accretion disk surrounding the MBH. The orange region at the
center represents the X-ray emitting region, probably the corona
structure above the accretion disk.
The absorption clouds
include both the BLR clouds outside of the accretion disk 
(yellow) and the dusty clumps (gray) in the dusty torus outside of
the broad line region (BLR). 
The spatial distribution of the clouds is
assumed to be axisymmetric.
The direction of the line of sight
(LOS), i.e., the viewing angle, is defined by the angle of
$\theta_{\LOS}$ with respect to the symmetric axis of the system ($z$),
and the LOS is on the $yz$ plane.
}
\label{fig:f1}
\end{figure}

Figure~\ref{fig:f1} shows a schematic picture for the spatial
distribution of numerous clouds rotating around the central engine, an
MBH-accretion disk system with enormous X-ray emission.  These clouds
are located in the two different regions, i.e., the BLR and the dusty
torus. The motions of those clouds are probably dominated by the
gravity of the central MBH (for the BLR clouds, see
\citealt{Gaskell88, Koratkar91, Sergeev99}; for the clumps in the
dusty torus, see \citealt{Elitzur06, Nenkova08}), and other effects,
such as the radiation pressure, on the cloud motion may be negligible.
For simplicity, we assume that all of those clouds and clumps are
spherical and are in Keplerian motion and on circular orbits around
the central MBH.\footnote{\citet{Bradley86} pointed out that the BLR clouds may
be on eccentric orbits as indicated by the non-Gaussian profile of
the emission lines. By alternatively assuming non-circular orbits, we find
no significant differences in the model results presented in this
paper. The effects, if any, on our model results due to the assumption
on the eccentricities of those clouds can be approximately compensated
by setting a slightly different radial distribution of the clouds.}.

To describe the motion of each cloud and the spatial distribution of
those clouds as a population, we use both an orthogonal
coordinate system ($x,y,z$) and a spherical coordinate system
($r,\theta,\phi$), with the origin located at the central MBH.
Here, $r$ is the distance to the central MBH, $\theta$
is the polar angle defined relative to the $z$-axis perpendicular to
the accretion disk, and $\phi$ is the azimuth angle (see
Fig.~\ref{fig:f1}). The spherical coordinate system is linked to the
orthogonal one by $(x,y,z)= (r\sin\theta \cos\phi, r\sin\theta
\sin\phi, r\cos\theta)$.  We set the distant observer to be on the
$yz$ plane with a direction of $(\theta,\phi)= (\theta_\LOS,\pi/2)$,
and the unit vector of this direction is $\hat{\mathbf{\ell}}= (0,
\sin \theta_\LOS, \cos\theta_\LOS)$ in the ($x,y,z$) coordinate system.

For a single cloud in a Keplerian motion, its orbit is determined by a
set of parameters $\mathbf{X}=(a_{\rm c},\theta_J,\phi_J)$, where
$a_{\rm c}$ are the semimajor axis of the orbit, and $\theta_J$ and
$\phi_J$ are the two angles defining the normal $\hat{n}_J$ of the orbital plane
with $\hat{n}_J$=$(\sin\theta_J \cos\phi_J, \sin\theta_J \sin\phi_J,
\cos\theta_J)$. Given the initial position of a cloud, its position at any
given moment can be obtained with that set of parameters. 
Each of those circular orbits can then be 
described by the set of parameters,
and the spatial distribution of those clouds as a population can be described
by a PDF $f_{\mathbf{X}}(\mathbf{X})$.
Assuming that the system is axisymmetric and the distribution of
$a_{\rm c}$ is independent of the distribution of the normal of the orbital
plane, the PDF can be further reduced to
$f_{a_{\rm c}} (a_{\rm c})|f_{\cos\theta_J}(\cos\theta_J)\sin\theta_J|/(2\pi)$.

\subsection{Eclipsing Events due to Individual
Clouds}~\label{subsec:ecp_event}

\begin{figure}
\centering
\includegraphics[scale=0.4]{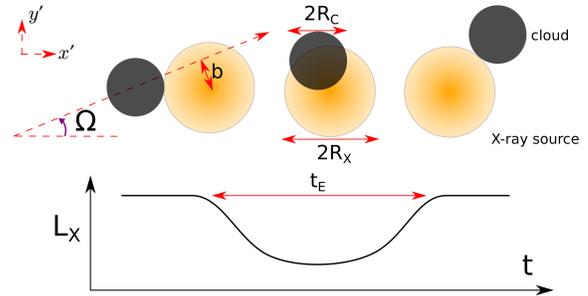}
\caption{ 
Schematic diagram for an X-ray eclipsing event by a cloud crossing
the LOS. The absorption cloud may be a cloud in the BLR region or a dusty clump,
as shown in Figure~\ref{fig:f1}.
The top row of this figure illustrates three phases of the
eclipse, i.e., the ingress phase (left), the greatest eclipse phase (middle),
and the egress phase (right). The observer is located at the direction of
pointing outward of the surface, with the $(x',y')$ coordinate system on the
surface.
The X-ray emitting region is assumed to
be spherical with a radius size of $R_{\rm X}$. The transiting direction on
the sky plane of the observer is determined by the impact parameter
$b$ and the direction angle $\Omega$ relative to a reference
direction (e.g., the horizontal line). The intervening cloud that
leads to the eclipse is also assumed to be spherical with a radius size of
$R_{\rm c}$. The duration of the eclipsing event is $t_{\rm E}$. The
bottom plot illustrates the variation of the X-ray flux due to the
eclipse.  
}
\label{fig:f2}
\end{figure}

If a cloud crosses the LOS and is in front of the central engine, then
an eclipse occurs and the cloud partially or completely blocks the
X-ray emitting region. In the image plane of the observer, the
trajectory of any eclipsing event can be described by the two parameters,
i.e., the impact parameter ($b$) and the eclipsing angle ($\Omega$),
as shown in Figure~\ref{fig:f2}. The eclipsing angle $\Omega$ is
defined as the angle between the motion direction of the cloud in the
image plane and a reference direction (the horizontal line from left
to right in Fig.~\ref{fig:f2}, i.e., the direction anti-parallel to
the $x$-axis in Fig.~\ref{fig:f1}), and it is given by 
\begin{eqnarray}
\cos \Omega & = & -\hat{v}_{yz} \cdot \hat{e}_{x}=-[ \hat{n}_J \times
(\hat{n}_J \times \hat{e}_x)] \cdot \hat{e}_{x} \nonumber \\ 
& = & \frac{\cos\theta_J}{|\cos\theta_J|}\sqrt{1-\sin^2 \theta_J
\cos^2 \phi_J},
\label{eq:cosomega}
\end{eqnarray}
and
\be
\sin \Omega = \sin \theta_J \cos \phi_J,
\label{eq:sinomega}
\ee
where $\hat{v}_{yz}$ is a unit vector with the same direction as the
velocity of a cloud when it crosses the $yz$ plane, and $\hat
{e}_x=(1,0,0)$ is a unit vector at the direction of the $x$-axis.  If
$\phi_J=\pi/2$ and $\theta_J \in [0, \pi/2)$, then $\Omega= 0$; if
$\phi_J=\pi/2$ and $\theta_J \in (\pi/2, \pi]$, then
$\Omega=\pi$. 
If $\theta_J=\pi/2$, then $\Omega$ is either $\pi/2-\phi_J$ or $\phi_J-\pi/2$.

The impact parameter $b$ is defined by 
\begin{eqnarray}
b & \equiv & a_{\rm c} (\hat{n}_J \cdot \hat{\mathbf{\ell}}) =  a_{\rm c}
\cos \omega    \nonumber \\ 
& = & a_{\rm c} (\sin \theta_J \sin \phi_J \sin \theta_\LOS + \cos
\theta_J \cos \theta_\LOS)
\label{eq:b}
\end{eqnarray} for $|b| \ll a_{\rm c}$ or $\cos \omega \ll 1$, which
is true for all of the cases considered in this paper since the size of
the central X-ray source is much smaller than the semimajor axes of
the absorption clouds. The $\omega$ in Equation (\ref{eq:b}) is the angle between the normal
of the orbital plane of a cloud and the LOS.  We have $b=0$ when
$\phi_J=\pi/2$ and $\theta_J = \pi/2 + \theta_\LOS$ (or more
generally, when $\sin \phi_J=-\cot\theta_J\cot\theta_\LOS$).  A cloud
crosses the LOS clockwise (or counter-clockwise) with respect to
the center of the X-ray source if $b<0$ (or $b>0$).

Assuming that the X-ray emitting region is spherical with a radius size of
$R_{\rm X}$ and a spherical cloud with a radius size of $R_{\rm c}$ crosses
the LOS, an eclipsing event occurs when $|b| < R_{\rm c} + R_{\rm X}$.
The following phases during this eclipsing event are illustrated
in Figure~\ref{fig:f2}: (1) an ingress phase when the
cloud starts to block the X-ray emission; (2) a maximum eclipsing
phase when the cloud blocks the projected X-ray emission region as much as
it can or even completely block the region, which leads to a dip in the
X-ray light curve; and (3) an egress phase when the part of the cloud
that first moved into the projected X-ray emission region starts to
move out. 

The duration of an eclipsing event caused by a cloud is
\be
t_{\rm E} \simeq 2  \left[ (R_{\rm X} + R_{\rm c})^2 - b^2\right]^{1/2} (a_{\rm
c}/GM_{\bullet})^{1/2},
\label{eq:tE}
\ee
where the cloud size $R_{\rm c}$ is assumed
to be a function of $a_{\rm c}$. If the eclipse event starts at a time $t_{\rm
s}$, the center of the cloud in the image plane is given by
\begin{eqnarray}
x'_{\rm c} & = & -b \sin \Omega + D_{\rm E} \left( \frac{t-t_{\rm
s}}{{t_{\rm E}}} -\frac{1}{2}\right) \cos \Omega, \\ y'_{\rm c} & = &
b \cos \Omega + D_{\rm E} \left( \frac{t-t_{\rm s}}{{t_{\rm E}}}
-\frac{1}{2}\right) \sin \Omega,
\end{eqnarray}
where $D_{\rm E} = 2\sqrt{(R_{\rm X}+R_{\rm c})^2-b^2}$, $t_{\rm s}< t
< t_{\rm s}+ t_{\rm E}$, and $t_{\rm s}$ is the start time of the
eclipse.

At any given time, the observational flux or luminosity curve is
obtained by
\be
L_{\rm X} \equiv \iint I(x',y') A_{\nu}(N_{\rm H}(R_{\rm c}, r_{x\rm
c})) dx'dy',
\ee 
where $I(x',y')$ is the intrinsic surface brightness of the X-ray emission
at a position $(x',y')$ on the image plane and $I(x',y')=0$ if
$x'^2+y'^2> R^2_{\rm X}$, $N_{\rm H}(R_{\rm c}, r_{\rm Xc})$ is the
projected column density at a distance $r_{\rm Xc}=\sqrt{[x'_{\rm
c}(t)-x']^2 + [y'_{\rm c}(t)-y']^2}$ from the cloud center on the
image plane, and $A_{\nu}$ represents the fraction of X-ray photons
penetrating through the cloud and received by the distant observer,
with $A_{\nu} = 1$ for the unblocked region with $x'^2+y'^2 \leq
R^2_{\rm X}$ and $A_{\nu}<1$ for the blocked region. ({\it Note here that the
X-ray eclipsing discussed in this paper is different from the cases
of planet transits, for which the emission from the area of a star
covered by a front planet is completely blocked, i.e., $A_{\nu} =0$.})
The ratio of the observational X-ray flux to the intrinsic one
(without eclipsing) is 
\be
R_{L_{\rm X}} \equiv \frac{L_{\rm X}}{L_{\rm X,0}} = \frac{\iint
I(x',y') A_{\nu}(N_{\rm H}(R_{\rm c}, r_{\rm Xc})) dx'dy'}{\iint
I(x',y') dx'dy'}.
\label{eq:R}
\ee

If the hydrogen number density ($n_{\rm H}$) of a cloud is uniform,
i.e., $n_{\rm H}$ is a constant within the cloud, then $N_{\rm
H}(R_{\rm c}, r_{\rm Xc}) = 2 n_{\rm H} \sqrt{R^2_{\rm c} - r^2_{\rm Xc}}$
for $r_{\rm Xc} \leq R_{\rm c}$ and $0$ for $r_{\rm Xc} > R_{\rm
c}$.  If the X-ray emissivity is also uniform, then $I(x',y') \propto
\sqrt{R^2_{\rm X} -x'^2-y'^2}$.  If the X-ray surface brightness is
homogeneous, then $I(x',y')$ is a constant, we have
\be
R_{L_{\rm X}}= \frac{\iint A_{\nu}(N_{\rm H}(R_{\rm c}, r_{\rm Xc}))
dx'dy'}{\pi R^2_{\rm X}}.
\label{eq:R2}
\ee
We define an ``effective column density'' as
\be
N_{\rm H, eff}\equiv A_{\nu}^{-1}(R_{L_{\rm X}}),
\label{eq:RAnu}
\ee
where $A_{\nu}^{-1}$ is the inverse function of $A_{\nu}$. The
effective column density may correspond to the absorption column
density directly measured from observations.

The bottom panel of Figure~\ref{fig:f2} illustrates the X-ray
variation due to an eclipsing event. The detailed calculations of the
X-ray variations at a given frequency range due to eclipsing events
are described in Section~\ref{sec:num_simu}.

For individual AGNs, we expect that a number of X-ray eclipsing events
can be detected over a substantially long observational period. It is
also possible that more than one cloud crosses the LOS at the same time
and covers (part of) the X-ray source, which lead to complex X-ray
variations (see Figure~\ref{fig:f6}). For such a case, $N_{\rm H,eff}$
at a given time is roughly the summation of the effective column
density due to each of the LOS crossing clouds.  The statistics of
those X-ray eclipsing events should depend on the spatial distribution
of the absorbing clouds in the AGN and the intrinsic properties of
those clouds.

\subsection{Event Rates of the X-ray Eclipses}

Given the total number of the clouds $N_{\rm tot}$ and their spatial
probability distribution function, the event rate of the X-ray eclipses caused
by the clouds with parameters $\mathbf{X}$ within the range
$\mathbf{X}\rightarrow\mathbf{X}+d\mathbf{X}$ can be given by
\begin{eqnarray}
& & d \mathscr{R} =  \frac{N_{\rm tot}f_{\mathbf{X}}(\mathbf{X})}{T_{\rm orb}(a_{\rm c})}d\mathbf{X} ,\nonumber \\
& = & da_{\rm c} d\cos\theta_J d\phi_J \frac{N_{\rm
tot}f_{a_{\rm c}}(a_{\rm c})}{2\pi T_{\rm orb}(a_{\rm c})}
f_{\cos\theta_J}(\cos\theta_J), \nonumber \\
   & = & - da_{\rm c} db d\Omega \frac{N_{\rm tot}f_{a_{\rm c}}(a_{\rm
c})}{2\pi T_{\rm orb}(a_{\rm c})} f_{\cos\theta_J}(\cos\theta_J) \sin \theta_J
\left| \frac{\partial (\theta_J, \phi_J)}{\partial (b,\Omega)}\right|,
\nonumber \\
\label{eq:eventrate}
\end{eqnarray}
where $T_{\rm orb}(a_{\rm c})=2\pi (a_{\rm
c}^3/G\bh)^{1/2}$ is the orbital period of a cloud with semimajor axis
$a_{\rm c}$, $\frac{\partial (\theta_J, \phi_J)}{\partial (b,\Omega)}$
is the Jacobian determinant, $\Omega$ and $b$ are given by
Equations~(\ref{eq:cosomega})--(\ref{eq:b}),
$\Omega \in (0, 2\pi]$, $b\in [-R_X-R_{\rm c}(a_{\rm
c}), R_X+R_{\rm c}(a_{\rm c})]$, $a_{\rm c}\in [a_{\rm c,min},a_{\rm
c,max}]$, and $a_{\rm c,min}$ and $a_{\rm c,max}$ are the smallest
and the largest values for the semimajor axes of those clouds,
respectively.

For most of the eclipsing clouds, we have $\cos \omega \simeq 0$ and $\omega
\simeq \pi/2$ since $b\ll a_{\rm c}$, and $\left| \frac{\partial
(\theta_J, \phi_J)}{\partial (b,\Omega)} \right| = \frac{\sin
\omega}{a_{\rm c}\sin \theta_J} \simeq \frac{1}{a_{\rm
c}\sin\theta_J}$. Therefore, we approximately have $\cos \theta_J
\simeq \sin \theta_{\LOS} \cos \Omega$ according to
Equations~(\ref{eq:sinomega}) and (\ref{eq:b}), and
the integration of Equation (\ref{eq:eventrate}) can be further reduced to 
\begin{eqnarray}
\mathscr{R} & = & \int d \mathscr{R} \nonumber \\
            & \simeq &  {N_{\rm tot}R_{\rm X}} \int^{\infty}_{0} \frac{
f_{a_{\rm c}}(a_{\rm c}) \left[1+R_{\rm c}(a_{\rm c})/R_{\rm X}\right]}{
a_{\rm c}T_{\rm orb}(a_{\rm c})} da_{\rm c} \nonumber \\ & &
\times \int^{2\pi}_0 \frac{f_{\cos\theta_J}(\sin\theta_\LOS \cos\Omega)}{\pi} d\Omega, \nonumber \\
& = & N_{\rm tot} \mathscr{H}(\theta_\LOS) \frac{(G\bh)^{1/2}
R_{\rm X}}{2\pi \left< a_{\rm E} \right>^{5/2}} \nonumber \\ & \simeq
& 1.4 {\rm day}^{-1} \mathscr{H}(\theta_\LOS)\left(
\frac{10^7\msun}{\bh} \right) \left( \frac{N_{\rm tot}}{10^7}\right)
\times \nonumber \\ & & \left( \frac{R_{\rm X}}{5r_{\rm g}}\right)
\left( \frac{\left<a_{\rm E}\right>}{10^4r_{\rm g}}\right)^{-5/2},
\label{eq:ER}
\end{eqnarray}
where $ \mathscr{H}(\theta_\LOS) \equiv \int^{2\pi}_0 d\Omega
f_{\cos\theta_J}(\sin\theta_{\LOS} \cos\Omega) / \pi $
is determined by the distribution of the angular momenta of the clouds
and the direction of the LOS. If $f_{\cos\theta_J}$ is a uniform distribution,
i.e., $f_{\cos\theta_J}=1/2$, then $\mathscr{H}(\theta_\LOS)=1$.
The effective semimajor axis of the eclipsing clouds $\left< a_{\rm E}
\right>$ is defined by 
\begin{equation}
\left< a_{\rm E}\right>^{-5/2} = \int^{\infty}_{0} da_{\rm c}
f_{a_{\rm c}}(a_{\rm c}) \left[1+R_{\rm c}(a_{\rm c})/R_{\rm X} \right]/
a^{5/2}_{\rm c},
\label{eq:aE}
\end{equation}
which is mainly determined by $f_{a_{\rm c}}(a_{\rm c})$. It may also depend on
the ratio of the size of the eclipsing clouds to the size of the X-ray
emitting region if the cloud size depends on the semimajor axis of the
cloud.

According to Equation~(\ref{eq:ER}), the distribution function of the
semimajor axis of those eclipsing clouds is given by
\be
f_{a_{\rm c},\rm E}(a_{\rm c})=\frac{\langle a_{\rm E}\rangle^{5/2}}{a_{\rm
c}^{5/2}}f_{a_{\rm c}}(a_{\rm c})\left[1+\frac{R_{\rm c}(a_{\rm c})}{R_{\rm
X}}\right].  
\label{eq:facE}
\ee 
This PDF indicates that the X-ray eclipsing events are preferentially
contributed by those clouds at relatively close distances to the
central MBH or those clouds with relatively large sizes (see also the
bottom-right panel in Figure~\ref{fig:f5} below). 

\subsection{Mean number of the clouds crossing the line of sight at a
given time} 

The mean number of the
clouds crossing the LOS at a given time is
\begin{eqnarray} 
\langle N \rangle_\LOS & = & \int t_{\rm E}d\mathscr{R}, \nonumber \\
 & \simeq & - \int^{\infty}_{0} da_{\rm c}
\int^{R_X+R_{\rm c}}_{-R_X-R_{\rm c}} db \nonumber \\ & &
\times \int^{2\pi}_0 d\Omega \frac{N_{\rm tot}f_{a_{\rm c}}(a_{\rm c}) t_{\rm
E}}{2\pi a_{\rm c} T_{\rm orb}(a_{\rm c})} f_{\cos\theta_J}(\sin \theta_\LOS
\cos\Omega)  \nonumber \\
& \simeq & 0.6 \mathscr{H}(\theta_\LOS)
\left(\frac{N_\tot}{10^7}\right) \left(\frac{R_X}{5 r_\g}\right)^2
\left(\frac{\langle a_N\rangle} {10^4 r_\g} \right)^{-2}, \nonumber \\
\label{eq:nlos}
\end{eqnarray}
where $t_{\rm E}$ in Equation (\ref{eq:tE}) is used and $\langle a_{\rm N}\rangle$ is defined through
\begin{equation}
\langle a_{\rm N}\rangle^{-2}=\int \left[1+ R_{\rm c}(a_{\rm
c})/R_X\right]^2 \left[f_{a_{\rm c}}(a_{\rm c})/a_{\rm c}^2 \right] da_{\rm c}.
\end{equation}
Note here that the dependence of $\langle N \rangle_\LOS$ on the
MBH mass is only through the dependence of $N_{\rm tot}$ or
$\mathscr{H}(\theta_\LOS)$ on the MBH mass, if the size of the X-ray source and $\left<a_{\rm N}\right>$ 
scale linearly with the gravitational radius of the central MBH (and the MBH
mass). According to Equation~(\ref{eq:nlos}), $\langle N\rangle_\LOS
\propto \langle a_{N}\rangle^{-2}$, the number of clouds crossing the
LOS depends on the size and spatial distribution of the clouds.
Similar to $\mathscr{R}$, $\langle N\rangle_\LOS$ also scales linearly with the
total number of clouds and depends on $\mathscr{H}(\theta_\LOS)$ (see
also the top-right panel of Figure~\ref{fig:f5} below).

\subsection{Statistical Properties of the Eclipsing Events} 

The trajectory of an eclipsing event on the sky plane is described by
the impact parameter $b$ and the eclipsing angle $\Omega$, and the
time duration of an eclipse is determined by $t_{\rm E}$.  The
statistical distribution of those parameters may be helpful for
understanding the X-ray variability resulting from the X-ray eclipsing
events.

\subsubsection{Probability distribution of the impact parameter}

The PDF of the impact parameter $|b|$ for an eclipse detected at any given time is given by 
\begin{eqnarray}
f_{|b|}(|b|) & = &
\frac{1}{\mathscr{R}}\frac{\partial}{\partial|b|}\int d\mathscr{R} \nonumber \\
& = &  -\frac{N_{\rm tot}}{\mathscr{R}}\frac{\partial}{\partial
|b|} \int^{\infty}_{0} d a_{\rm c} \frac{ f_{a_{\rm c}}(a_{\rm c}) }{ a_{\rm
c}T_{\rm orb}(a_{\rm c})} \nonumber \\ & & \times \int^{R_X+R_{\rm
c}(a_{\rm c})}_{|b|} d|b|' \int^{2\pi}_0 \frac{f_{\cos\theta_J}(\sin\theta_\LOS
\cos\Omega)}{\pi} d\Omega \nonumber \\
& = & \frac{1}{R_X} \frac{\int^{\infty}_{a_{\rm c,l}} f_{a_{\rm c}}(a_{\rm
c})a_{\rm c}^{-5/2} da_{\rm c}} {\langle a_{\rm E}\rangle^{-5/2}},
\label{eq:fb}
\end{eqnarray}
where $a_{\rm c,l}= \max[R^{-1}_{\rm c}(|b|-R_X),a_{\rm c,min}]$ if
$|b|>R_X$, $a_{\rm c,l}= a_{\rm c,min}$ if $|b|\leq R_X$, and
$R^{-1}_{\rm c} (|b|-R_X)$ is an inverse function of $R_{\rm c}(a_{\rm
c})= |b|-R_X$.  If all of the clouds have the same size, then
$f_{|b|}(|b|)=\frac{1}{R_X+R_{\rm c}}$ for $|b|\leq R_X+R_{\rm c}$.
For more general cases, such as those models listed in
Table~\ref{tab:t1}, $f_{|b|}(|b|)$ is a constant when $|b|\leq
R_X+R_{\rm c,min}$, and decreases with increasing $|b|$ when
$|b|>R_X+R_{\rm c,min}$, where $R_{\rm c,min}$ is the minimum radius size of
the clouds (see more on size distribution in Section~\ref{subsec:cloudproperies}).
 
\subsubsection{Probability distribution of the eclipse angle}

For any given LOS, the PDF of $\Omega$ is given by 
\be 
f_{\Omega}(\Omega) \simeq \frac{f_{\cos\theta_J}(\sin \theta_\LOS \cos \Omega)}{\pi
\mathscr{H}(\theta_\LOS)}.
\label{eq:fOmega} 
\ee 
As seen from equation~(\ref{eq:fOmega}), the PDF $f_{\Omega}$ is
mainly determined by the spatial distribution of those eclipsing
clouds (or their parent population).

We note here that, in principle,
the profile of the Fe K$\alpha$ line emission, if any, from the inner
disk of an AGN depends on the disk structure. If the eclipsing events
frequently occur with different $(b,\Omega)$, the changes of the Fe
K$\alpha$ profile during those events are determined by both the
parameters $(b,\Omega)$ \citep{Risaliti11} and the inner disk
structure (or the emissivity law of the line).  Therefore, the
variation of the Fe K$\alpha$ line emission due to eclipsing may also be
used to probe the kinematic structure of the parent population of
the eclipsing clouds and the inner disk structure. 

\subsubsection{Probability Distribution of the Eclipse Timescale}

The PDF of the eclipse timescale $t_{\rm E}$ can be obtained by
\begin{eqnarray}
 & & f_{t_{\rm E}}(t_{\rm E})  =  \frac{1}{\mathscr{R}}\frac{\partial}{\partial t_{\rm E}}\int d\mathscr{R}, \nonumber \\
& = & \frac{N_{\rm tot}}{E}\frac{\partial}{\partial
t_{\rm E}} \int^{\infty}_{0} d a_{\rm c} \frac{ f_{a_{\rm c}}(a_{\rm c}) }{
a_{\rm c}T_{\rm orb}(a_{\rm c})} \nonumber \\
& & \times \int^{R_X+R_{\rm
c}(a_{\rm c})}_{\sqrt{[R_X+R_{\rm c}(a_{\rm c})]^2-\frac{G\bh t_{\rm
E}^2}{4a_{\rm c}}}} db' \int^{2\pi}_0 \frac{f_J(\sin\theta_\LOS
\cos\Omega)}{\pi} d\Omega, \nonumber \\
& = & \int^\infty_0 \frac{\langle a_{\rm E}\rangle^{5/2}}{a_{\rm
c}^{5/2}} \left[1+\frac{R_{\rm c}(a_{\rm c})}{R_{\rm X}}\right]
\frac{t_{\rm E}}{t'_{\rm E}(a_{\rm c})} \frac{f_{a_{\rm c}}(a_{\rm c})da_{\rm
c}}{\sqrt{t'_{\rm E}(a_{\rm c})^2-t_{\rm E}^2}}, \nonumber \\
\label{eq:pdte}
\end{eqnarray}
where
\be
t'_{\rm E}(a_{\rm c})=2[R_{\rm X}+R_{\rm c}(a_{\rm
c})]/\sqrt{G\bh/a_{\rm c}}
\label{eq:te}
\ee
is the maximum eclipse duration (when $b=0$). 
The mean duration of the eclipsing events is given by $\langle t_{\rm
E} \rangle=\int t_{\rm E}f_{t_{\rm E}}(t_{\rm E})dt_{\rm E}=\langle N
\rangle_\LOS/ \mathscr{R}$, which can be further reduced to
\begin{eqnarray}
\langle t_{\rm E} \rangle &= & 0.45~{\rm day}
\left(\frac{\bh}{10^7\msun}\right) \left( \frac{R_X}{5r_g} \right)
\left(\frac{\langle a_{E}\rangle}{10^4r_g}\right)^{5/2}
\left(\frac{\langle a_{N}\rangle}{10^4r_g}\right)^{-2}. \nonumber \\
\label{eq:te_avg}
\end{eqnarray}
According to Equation (\ref{eq:te_avg}), $\langle t_{\rm E} \rangle
\propto \bh$ if $R_X$, $\langle a_{\rm E}\rangle$ and $\langle a_N
\rangle$ all scale linearly with the mass of the central MBH.
The typical
eclipsing duration for AGNs with central MBH mass in the range from
$10^6\msun$ to $10^9\msun$ is $\langle t_{\rm E} \rangle \sim 0.05$ to
$\sim 50$\,days, which can be monitored by most currently available
X-ray missions.  Note that the eclipsing time $t_{\rm E}$ due to
individual clouds can be as long as $10^3$\,days for those clouds
located at the outer part of the torus, or as short as $10^{-1}$\,days
for clouds located at the inner part of the BLR (see the bottom-left
panel of Figure~\ref{fig:f5}).


\section{Monte-Carlo Simulations of the X-ray Eclipses}
\label{sec:num_simu}

In this section, we use Monte-Carlo simulations to realize the
kinematical motion of the clouds in AGNs and the X-ray eclipses.  We adopt
eight models for the spatial distribution and the intrinsic properties
of the clouds as listed in Table~\ref{tab:t1}.  Some settings for the
parameters involved in the cloud model are well motivated by
observations as described below. The details of the Monte-Carlo
simulations are described in Section~\ref{subsec:directmodeling} and
the simulation results are presented in Section~\ref{subsec:results}. 

\subsection{Settings on the Spatial Distribution of the Clouds}
 \label{subsec:spatial}

Observations have shown that different broad emission lines are
originated from gas clouds or clumps at different distances from the
central illuminating source. Those line-emitting clouds and clumps are
distributed over a large spatial extent, from the inner edge of the
BLR to the outer region of the dusty torus. The high ionization
emission lines (C{\small IV}, He{\small II}, etc.) are originated from
a region closer to the central MBH than the low ionization emission
lines (H{\small $\beta$}, He{\small I}, etc.)
\citep[e.g.,][]{Gaskell86, Kaspi00, Peterson04, Landt08, Bentz10,
Pancoast11}. Some lines in the infrared (IR), e.g., O{\small I},
Pa$\epsilon$, Br$\gamma$, are emitted from a region even farther away,
extending to the dust sublimation radius $r_{\rm d}$, the presumed
inner edge of the dusty torus \citep{Landt08, Markowitz14}.  It was
also suggested that the torus may be a smooth continuation of the BLR
\citep{Schartmann05, Suganuma06, Elitzur06}. We thus assume that the
radial distribution of those clouds and clumps can be described by a
simple power-law function, and we will not analyze them separately in this
paper.

Infrared reverberation mapping observations have shown that the dust
sublimation radius is roughly given by
\begin{eqnarray}
r_{\rm d} &\simeq & 0.40{\rm pc} \left( \frac{L_{\rm bol}}{10^{45}{\rm
erg\,s}^{-1}}\right)^{1/2}\left(\frac{T_{\rm sub}}{1500\,{\rm
K}}\right)^{2.6} \nonumber \\ 
&\simeq & 0.45{\rm pc} \lambda^{1/2}_{\rm Edd}
\left(\frac{\bh}{10^7\msun}\right)^{1/2} \left( \frac{T_{\rm
sub}}{1500{\rm \,K}}\right)^{2.6},
\end{eqnarray}
where $L_{\rm bol}$ is the AGN bolometric luminosity, $\lambda_{\rm
Edd}\equiv L_{\rm bol}/L_{\rm Edd}$ is the Eddington ratio, $L_{\rm
Edd}\simeq 1.3 \times 10^{45} {\rm erg\,s}^{-1} (\bh/10^7\msun)$ is
the Eddington luminosity, and $T_{\rm sub}$ is the dust sublimation
temperature \citep{Suganuma06, Nenkova08}.  For all of the models listed
in Table~\ref{tab:t1}, we assume that $T_{\rm sub}=1500$\,K and
$\lambda_{\rm Edd}= 0.1$, since observations suggest that most AGNs (and
QSOs) accrete material via a rate close to $0.1-0.3$
\citep[e.g.,][]{Kollmeier06, Shen08}. If $\bh=10^7\msun$, $r_{\rm d}$
is approximately $3\times 10^5\rg$.

We assume that the radial distribution of those clouds (and clumps)
follows a broken power law with a transition radius $r_{\rm t}$, i.e., 
\be
f_{a_{\rm c}}(a_{\rm c}) \propto \begin{cases} \left(a_{\rm c}/r_{\rm t}
\right)^{\alpha_{a_{\rm c},1}}, & \quad \text{for } a_{\rm c} \leq r_{\rm t}, \\
\left(a_{\rm c}/r_{\rm t} \right)^{\alpha_{a_{\rm c},2}}, & \quad \text{for }
a_{\rm c} > r_{\rm t}.  \end{cases} 
\label{eq:fa}
\ee
The slopes $\alpha_{a_{\rm c},1}$ and $\alpha_{a_{\rm c},2}$ control the radial distribution
of those clouds. For example, for a lower $\alpha_{a_{\rm c},1}$, relatively
more clouds are located in the inner region; and for a larger $\alpha_{a_{\rm c},2}$,
relatively more clouds are located in the outer region. The transition
radius $r_{\rm t}$ may be proportional to the dust sublimation radius
$r_{\rm d}$, roughly the boundary between the BLR and the dusty torus.
The reason is that the gas clumps (or clouds) in the region outside of
the sublimation radius may be significantly less affected by the
radiation from the central source than those clouds within that
radius.  For all of the models in Table~\ref{tab:t1}, we assume that
$r_{\rm t} \sim 0.1 r_{\rm d}$.  In models C1-C4, the clouds are
more concentrated in the inner region compared with those in 
models A1, A2, B1, and B2; the clouds are more concentrated in the
outer region in models B1 and B2 compared with those in the other
models.  How the model results depend on the parameter settings, i.e.,
the inner and the outer boundaries for the spatial distribution of
clouds, the Eddington ratio, and the transition radius $r_{\rm t}$ is
discussed in Section~\ref{sec:pds}.

The PDF of the orbital planes (or the direction of the orbital angular momenta)
of those clouds over the solid angle is denoted as
$f_{\cos\theta_J}(\cos\theta_J)/2\pi$, i.e., the fraction per solid
angle ($|d\cos\theta_J d\phi_J|$).
If the PDF of $f_{\cos\theta_J}(\cos\theta_J)$ is uniform, then
$f_{\cos\theta_J}(\cos\theta_J) = 1/2$. However, the spatial distribution of the clouds
is probably flattened
\citep[e.g,][]{Bentz10, Li13}.  Considering this flattening, we assume
that the distribution of $\theta_J$ follows a Schwarzschild or Rayleigh
distribution with
\be
f_{\cos\theta_J}(\cos \theta_J) d\cos\theta_J\propto
 \theta_J e^{-\frac{\theta_J^2}{2\sigma^2_{\theta_J}}}d\theta_J,
\label{eq:fcostheta1}
\ee
where we set $\sigma_{\theta_J}=\pi/9$ for all of the models listed in
Table~\ref{tab:t1}. Considering that the distribution in Equation
(\ref{eq:fcostheta1}) was derived with a small $\theta_J$ approximation, we
also try the following distribution
\be
f_{\cos\theta_J}(\cos \theta_J) d\cos\theta_J\propto
\sin\theta_J e^{-\frac{\sin^2\theta_J}{2\sigma^2_{\sin\theta_J}}}d\theta_J,
\label{eq:fcostheta2}
\ee
where $\sigma_{\sin\theta_J}=\sin(\pi/9)$. 
The total angular momentum of the clouds is none-zero in the distribution of
Equation (\ref{eq:fcostheta1}), and zero in Equation (\ref{eq:fcostheta2}),
which represents some extreme case of the kinematic distribution of the
clouds.
We find that choosing a different $\sigma_{\theta_J}$ or a different
distribution function
of $f_{\cos\theta_J} (\cos\theta_J)$ affects only the function
$\mathscr{H}(\theta_\LOS)$, and thus $\mathscr{R}$ and $\langle
N\rangle_\LOS$ (which can be obtained straightforwardly from
Equations~\ref{eq:ER} and \ref{eq:nlos}), but not the shape of the PSDs.
For simplicity, we only present the results obtained by using Equation
(\ref{eq:fcostheta1}).

\begin{deluxetable*}{c|ccc|cc|cc|ccc|ccc}
\tablewidth{18.5cm}
\rotate
\tabletypesize{\scriptsize}
\tablecaption{Parameters in the cloud models}
\tablecolumns{2}
\startdata
\tablehead{
\colhead{Model} & \colhead{$\alpha_{a_{\rm c},1}$} &
\colhead{$\alpha_{a_{\rm c},2}$}  & \colhead{$a_{\rm
c,min}$\tablenotemark{a}} & \colhead{$R_{\rm c,0}$} &
\colhead{$\alpha_{R_{\rm c}}$\tablenotemark{b}} & \colhead{$n_{\rm H,0}$}
& \colhead{$\alpha_{n_{\rm H}}$\tablenotemark{c}} & \colhead{$\beta_{\rm h}$} &
\colhead{$\beta_{\rm l}$} & \colhead{$T_{\rm B}^{N_{\rm H}}$
(day)\tablenotemark{d}} & \colhead{$\gamma_{\rm h}$} &
\colhead{$\gamma_{\rm l}$} & \colhead{$T_{\rm B}^{L_{\rm X}}$
(day)\tablenotemark{e}} 
} 
%
%
A1&   1 & -0.5  & $10^3$ & 2   & 1   & 1  & -1.5  & $-4.0\pm 0.3$ &$
-0.94\pm 0.08$ &$ 2.1\pm0.6$ &$ -3.9\pm 0.2$ &$ -1.0\pm 0.1$ &$
2.2\pm0.6$\\
A2&   1 & -0.5  & $10^3$ & 2   & 1   & 1  & -1     & $-4.0\pm0.1$ &$
-1.5\pm0.1$ &$ 2.5\pm0.7$ &$ -3.7\pm0.2$ &$ -1.4\pm0.1$ &$
2.5\pm0.9$\\
B1&   1 & 0      & $10^3$ & 2   & 1   & 1  & -1.5  & $-4.2\pm 0.2$ &$
-1.0\pm0.1$ &$ 2.3\pm0.7$ &$ -4.0\pm0.4$ &$ -1.1\pm0.1$ &$
2.7\pm0.9$\\
B2&   1 & 0      & $10^2$ & 0.2 & 1  & 60 & -1.5  & $-4.2\pm0.2$ &$
-1.12\pm0.03$ &$ 1.6\pm0.3$ &$ -4.0\pm0.2$ &$-1.17\pm0.04$ &
$ 1.8\pm0.4$\\
C1&   0 & -0.5  & $10^3$ & 2   & 1   & 1  & -1.5  & $-6.3\pm0.2$ &$
-0.88\pm0.02$ &$ 0.41\pm0.09$ &$ -6.4\pm0.2$ &$ -0.98\pm0.03$ &$
0.4\pm0.3$\\
C2&   0 & -0.5  & $10^2$ & 0.2 & 1  & 60 & -1.5  & $-2.8\pm0.1$ &$
-0.81\pm0.08$ &$1.8\pm0.3$ &$ -2.6\pm0.2$ &$ -0.91\pm0.03$ &$ 1.6\pm0.3$\\
C3&   0 & -0.5  & $10^2$ & 0.2 & 1  & $10^2$ & -2    & $-2.3\pm0.1$ &$
-0.20\pm0.04$ &$ 1.6\pm0.4$ &$ -2.11\pm0.09$ &$-0.18\pm0.04$ &$ 1.7\pm0.5$\\
C4&   0 & -0.5  & $10^2$ & 0.6 &0.5 & 60 & -1.5 & $-2.56\pm0.09$ &$
0.39\pm0.03$ &$ 0.4\pm0.1$ &$ -2.7\pm0.2$ &$ 0.33\pm0.04$ &$ 0.27\pm0.08$ \\
\enddata
\label{tab:t1}
\tabletypesize{\small}
\tablenotetext{a}{The $\alpha_{a_{\rm c},1}$ and $\alpha_{a_{\rm c},2}$ are the power laws in the radial
distribution of the clouds with low semimajor axes and high semimajor axes,
respectively; and $a_{\rm c,min}$ is the minimum semimajor axis of the clouds, in units of $r_\g$. See Equation (\ref{eq:fa}).}
\tablenotetext{b}{The $\alpha_{R_{\rm c}}$ is the power law in the size distribution of
the clouds, and $R_{\rm c,0}$ is the size of the clouds closest to the
central MBH (with $a_{\rm c}=a_{\rm c,min}$), in units of $r_\g$. See
Equation (\ref{eq:Rc}).}
\tablenotetext{c}{The $\alpha_{n_{\rm H}}$ is the power law in the hydrogen
density
distribution of the clouds, and $n_{\rm H,0}$ is the density of those
individual clouds with $a_{\rm c}= a_{\rm c,min}$, in units of
$10^{11}~{\rm cm}^{-3}$.  See Equation (\ref{eq:nH}).}
\tablenotetext{d}{The parameters $\beta_{\rm l}$, $\beta_{\rm h}$, and
$T_{\rm B}^{N_{\rm H}}$ are the PSD slope at low frequencies, the PSD
slope at high frequencies, and the break timescale $(=1/\nu_{\rm
B})$ obtained from the best-fit to the PSD of the mock column density
variations by using Equation~(\ref{eq:bkpow}). For all the models listed in the table, the viewing angle
is set to $60\arcdeg$, and $\bh=10^7\msun$.  Our calculations also show that the shape of
the PSD does not depend on the viewing angle. The break timescale $T_{\rm B}$
may correlate with $\bh$, as described in Section~\ref{subsec:relationship}. }
\tablenotetext{e}{The parameters $\gamma_{\rm l}$, $\gamma_{\rm h}$, and
$T_{\rm B}^{L_{\rm X}}$ are the slope at low frequencies, the slope at
high frequencies, and the break timescale $(=1/\nu_{\rm B})$
obtained from the best-fit to the PSD of the mock $2-10$\,keV flux
variations by using Equation~(\ref{eq:bkpow}).
}
\end{deluxetable*}

\subsection{$N_\tot$}

The total number of the clouds in the BLR and the dusty torus can be
estimated for some sources according to the fluctuations in the emission
line profiles caused by a finite number of discrete line emitters.
For example, \citet{Dietrich99} estimated that the total number of
the clouds in the BLR of 3C\,273 is $\sim 10^8$;  \citet{Arav97} and
\citet{Laor06} put a lower limit of $3\times 10^6$ and $\sim
10^4-10^5$ on the number of BLR clouds for Mrk\,335 and NGC\,4395,
respectively; \citet{Arav98} found that the total number of the clouds in
the BLR of NGC\,4151 has to be $\sim 3\times 10^7-10^8$ in order to
generate the observed profile of the emission lines. However, it is
not clear whether the total number of the clouds in the BLR and the dusty
torus depends on AGN properties, e.g., the MBH mass, the Eddington
ratio.  For simplicity, we assume that the total number of clouds
is $N_\tot=10^7$ for those AGN models investigated in this paper. We
have also further checked that choosing a different $N_{\rm tot}$,
say, $10^6$ or $10^8$, does affect our estimates on the eclipsing
event rate ($\mathscr{R}$) and the mean number of clouds crossing the LOS at any
given time ($\langle N \rangle_\LOS$), but does not affect our results
on the shape of the PSD presented in Section~\ref{sec:pds}.

\subsection{Settings on the Intrinsic Properties of the Clouds}
\label{subsec:cloudproperies}

The detected X-ray eclipsing events in some AGNs so far suggest that
the sizes of the eclipsing clouds should be on the order of $\sim
1-10^2 \rg$ \citep[e.g.,][]{Sanfrutos13,Markowitz14}. It also appears
that the sizes of the eclipsing clouds increase with their increasing
distances from the central engine \citep{Markowitz14}.  However,
the exact distribution of the sizes of the eclipsing clouds and their
parent population is not currently known. For simplicity, we assume
that the radius size of a cloud depends on its semimajor axis, i.e.,
\begin{equation}
R_{\rm c}(a_{\rm c}) = R_{\rm c,0} (a_{\rm c}/a_{\rm c,min})^{\alpha_{R_{\rm c}}},
\label{eq:Rc}
\end{equation}
where $R_{\rm c,0}$ is the size of those clouds located at the inner edge of the
BLR ($a_{\rm c,min}$).  As listed in Table~\ref{tab:t1}, we set
$(\alpha_{R_{\rm c}},R_{\rm c,0}) =(1,2r_\g)$ for the models A1, A2, B1, and C1;
$(\alpha_{R_{\rm c}},R_{\rm c,0}) =(1,0.2r_\g)$ for the models B2, C2, and C3;
and $(\alpha_{R_{\rm c}},R_{\rm c,0}) = (0.5, 0.6r_\g)$ for the model C4. With these
settings, the clouds at a distance of $\simeq10^3r_\g$ from the
central engine have almost the same size ($\simeq 2r_\g$) in all of these
models; the filling factor of the clouds within the outer boundary
$\la 1.4$\,pc (for $\bh=10^7\msun$) is much smaller than
one; and the covering factor is in the range of $0.5-0.8$, compatible
with the constraints obtained from infrared observations of the torus
\citep[e.g.,][]{Ichikawa15, Nenkova08}. 

Note here that the possible collisions among the clouds are ignored in this
study. In reality, those clouds in the BLR and the dusty torus may be
on eccentric orbits and their velocities disperse, so that they may collide
with each other and be destroyed. The collision rate can be
roughly estimated by $\sim N_{\text{\tot}} \Sigma \Delta v /V $,
where $\Sigma\sim \pi \langle R^2_{\text{c}}\rangle$ is the mean cross
section of those clouds, $\langle R^2_{\text{c}}\rangle$ is the mean of the square radius of those
clouds, $\Delta v$ is the mean velocity dispersion, and $V$ is the
volume. If $\Delta v$ is on the order of the Keplerian velocity, we
find that the collision rate is roughly a few times of $10^{-5}$\,per
year or less per cloud, for $N_{\text{tot}}=10^7$ and the size
distribution of the clouds listed in Table~\ref{tab:t1}, which
verifies the validity of ignoring collisions in a period of less
than a hundred years investigated in this study.

We assume that the hydrogen density in a single cloud is uniform and
it depends on the distance of the cloud from the central engine.
Observations suggest that the densities of BLR clouds correlate with
the FWHM of the broad emission lines and the typical hydrogen densities of BLR
clouds are in the range of $10^{8}-10^{11}{\rm cm}^{-3}$ or even bigger
(e.g., \citealt{Peterson97,Osterbrock06}).  Therefore, we assume that the hydrogen
densities of the clouds follow a power-law function of their distance to
the central MBH, i.e.,
\be
n_{\rm H}=n_{\rm H,0} (a_{\rm c}/ a_{\rm c,min})^{\alpha_{n_{\rm H}}},
\label{eq:nH}
\ee
where $n_{\rm H,0}$ is the hydrogen density of
those clouds that are the closest to the central MBH with $a_{\rm
c}=a_{\rm c,min}$.  The value of the power-law index is in the range
of $-2<\alpha_{n_{\rm H}}<0$. For the models listed in Table~\ref{tab:t1}, we
set $\alpha_{n_{\rm H}}=-1$ for model A2, $\alpha=-2$ for model C3, and
$\alpha=-1.5$ for all of the other models.   We set the density
$n_{\rm H,0}=6\times10^{11}{\rm cm}^{-3}$ at $a_{\rm c,min}=10^2r_\g$
for models B2, C2, and C4, and $n_{\rm H,0}=10^{11}{\rm cm}^{-3}$
at $a_{\rm c,min}=10^3 r_\g$ for all of the other models. According to
these settings, we have $n_{\rm H}\simeq 10^{11}{\rm cm}^{-3}$ at
$a_{\rm c}=10^3 r_\g$ for all of the models in Table~\ref{tab:t1}.

\subsection{Size of the X-ray Emitting Region}

Observations have suggested that the X-ray emitting region of AGNs is
close to the central MBH and its size is small. For example,
\citet{Dai10} find that the sizes of the X-ray emitting regions in
some lensed QSOs are smaller than $10r_\g$ by using the microlensing
technique to map the structure of the accretion disks around those
QSOs; \citet{Markowitz14} find a similar size for the X-ray emitting
region by using individual X-ray eclipsing events \citep[see also][]{
Lamer03b, Sanfrutos13}.  For simplicity, we set $R_X=5r_\g$ for all
the models in this paper. In principle, the sizes of the emitting
regions in different AGNs can be somewhat different from each other.
The effects of the different choices of  $R_{\rm X}$ on our results 
are discussed in Section~\ref{sec:pds}. 

For simplicity, we also assume that the brightness distribution of the
X-ray source is homogeneous. If alternatively we assume that the
emissivity is uniform, we find no significant difference in the model
results presented in this paper.

\begin{figure*}
\centering
\includegraphics[scale=0.8]{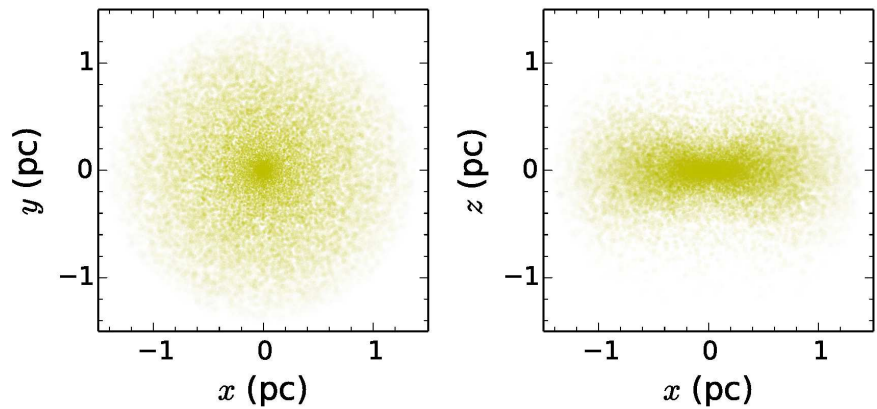}
\caption{Illustration map for the spatial distribution of those clouds
resulting from model C1 (see Table~\ref{tab:t1}). The left and right panels 
show the face-on
view and the edge-on view, respectively.  }
\label{fig:f3}
\end{figure*}

\begin{figure*}
\centering 
\includegraphics[scale=1.0]{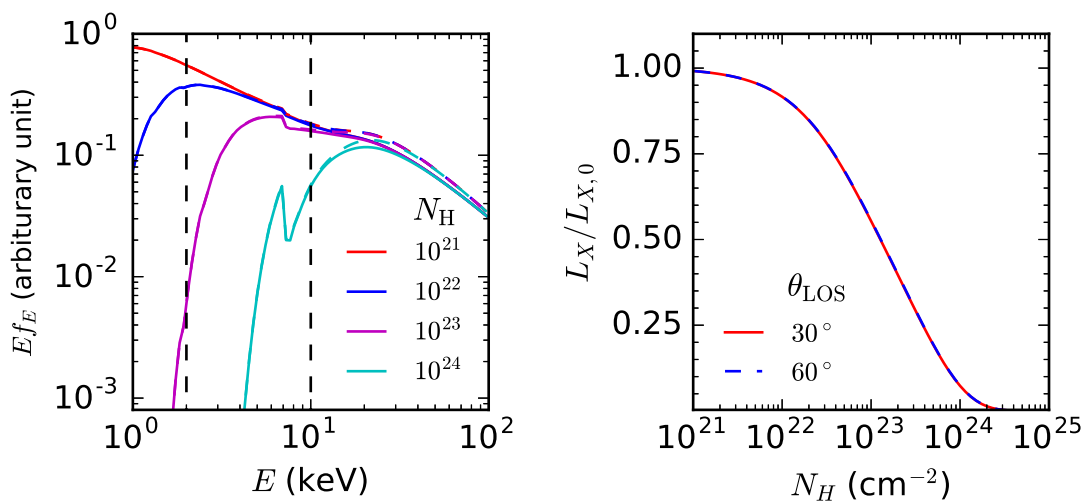}
\caption{Left panel: the mock observational X-ray spectra resulting
from different absorption column densities. The red, blue, magenta,
and cyan lines represent the results for an absorption of $N_{\rm
H}=10^{21}$, $10^{22}$, $10^{23}$, and $10^{24}$\,cm$^{-2}$,
respectively. The dashed and the solid lines with the same color
represent the results for the case with a view angle of
$\theta_\LOS=60\arcdeg$ and $30\arcdeg$, respectively. Here the view
angle is relevant because it needs to be set in the model $\rm pexrav$
for generating the K$\alpha$ emission line and the reflection
component. Right panel: the ratio of the mock observed flux ($L_X$) to
the intrinsic X-ray flux ($L_{X,0}$) at the $2-10$\,keV band as a
function of the absorption column density obtained from the model
phabs*pexrav in the Xspec package.  
The red solid and the blue dashed lines show the results
for the cases with $\theta_\LOS=30\arcdeg$ and $60\arcdeg$,
respectively.  
See
Section~\ref{subsec:directmodeling} for the details of the model
parameters.
}
\label{fig:f4}
\end{figure*}
%


\subsection{Variations of the Absorption Column Density and the X-ray
Emission}
\label{subsec:directmodeling}

With the above settings, we perform Monte-Carlo simulations to follow
the Keplerian motions of the parent population of those eclipsing
clouds with random initial orbital true anomalies. 
At any given moment,
the positions of all of the clouds on the sky plane of the observer (with
a view angle of $\theta_\LOS$) can be obtained. As an example,
Figure~\ref{fig:f3} shows the spatial distribution of the clouds generated
from model C1, in which the clouds are centrally concentrated
because of $\alpha_{n_{\rm H}}<0$. All of the clouds that happen to transit the X-ray
emitting region and lead to X-ray eclipses can be identified at any
given moment for each model. The changes of the absorption column
densities and the X-ray flux with time can then be obtained according
to the description in Section~\ref{subsec:ecp_event} and the
procedures below. In the meantime, we can also obtain the total number
of clouds across the LOS ($\langle N\rangle_\LOS$; see
Eq.~(\ref{eq:nlos})) and the event rate of eclipses ($\mathscr{R}$); see
Eq.~(\ref{eq:ER})).

\begin{figure*}
\centering
\includegraphics[scale=0.8]{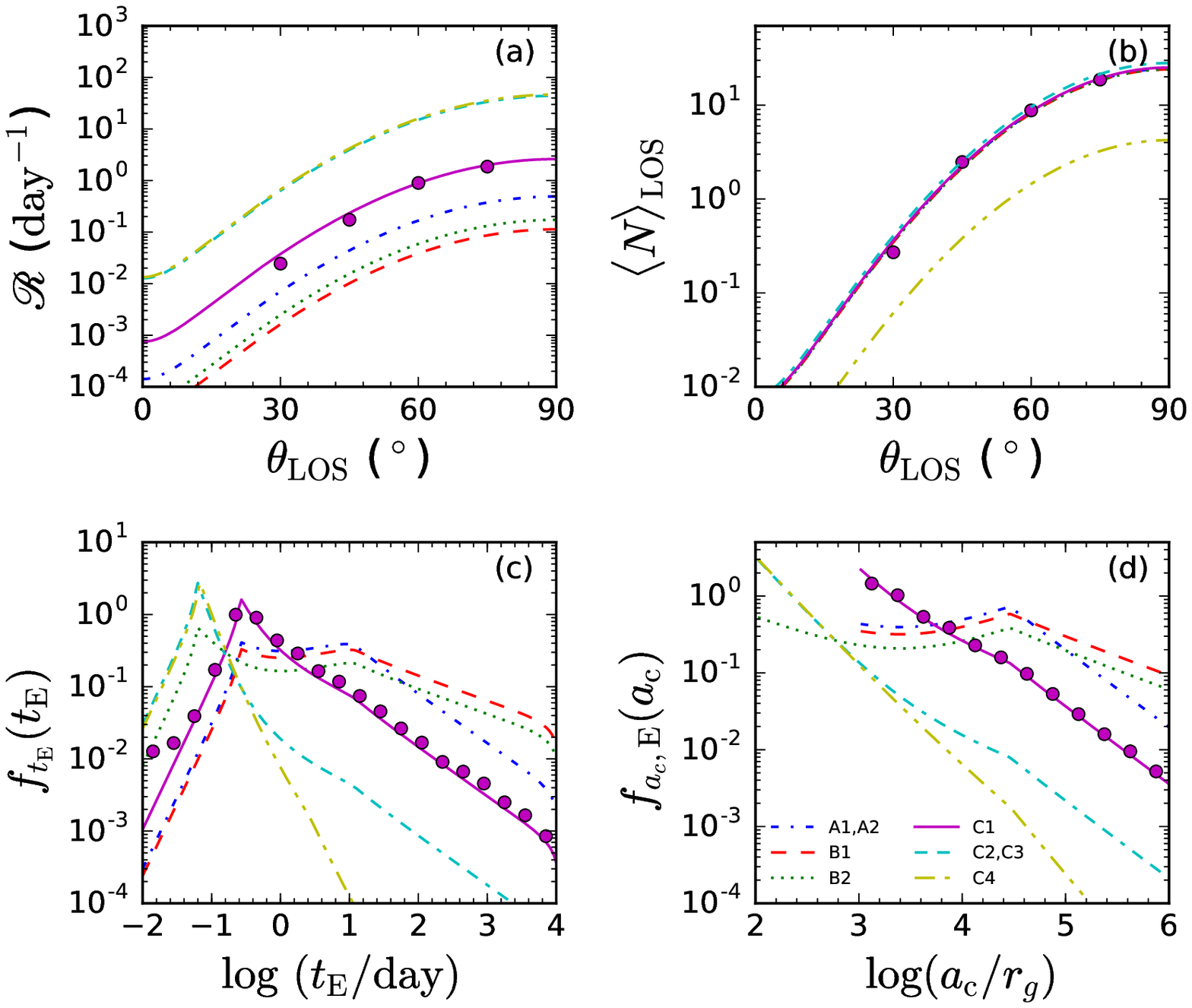}
\caption{Statistical properties of the X-ray eclipses resulting from
those models listed in Table~\ref{tab:t1}. Panels (a) and (b) show the
event rate of the X-ray eclipses ($\mathscr{R}$) and the time-averaged
number of the eclipses ($\langle N \rangle_\LOS$) as a function of the
viewing angle $\theta_\LOS$, respectively. Panels (c) and (d) show the
PDFs for the time of eclipse ($f_{t_{\rm E}}(t_{\rm E})$) and the semimajor axes
of the eclipsing clouds ($f_{a_{\rm c},\rm E}(a_{\rm c})$; Eq.~\ref{eq:facE}), respectively. The different
colors in the panels represent the results from the different models, as
indicated by the text in panel (d). 
In each panel, the lines
represent the results obtained from the analytical estimations
presented in Section~\ref{sec:analytic}, and the magenta circles
represent the results obtained from the Monte-Carlo simulations described
in Section~\ref{sec:num_simu}. For view clarity, we only show the
Monte-Carlo simulation results for one model (C1), as the Monte-Carlo
simulation results are well consistent with the analytical
estimations. Since $f_{t_{\rm E}}(t_{\rm E})$ and $f_{a_{\rm c},\rm E}(a_{\rm c})$ are independent
of the view angle $\theta_\LOS$ for all the models, here we only show
the results for those models obtained by assuming $\theta_\LOS=60\arcdeg$ in
panels (c) and (d). }
\label{fig:f5}
\end{figure*}

\begin{figure*}
\centering
\includegraphics[scale=0.7]{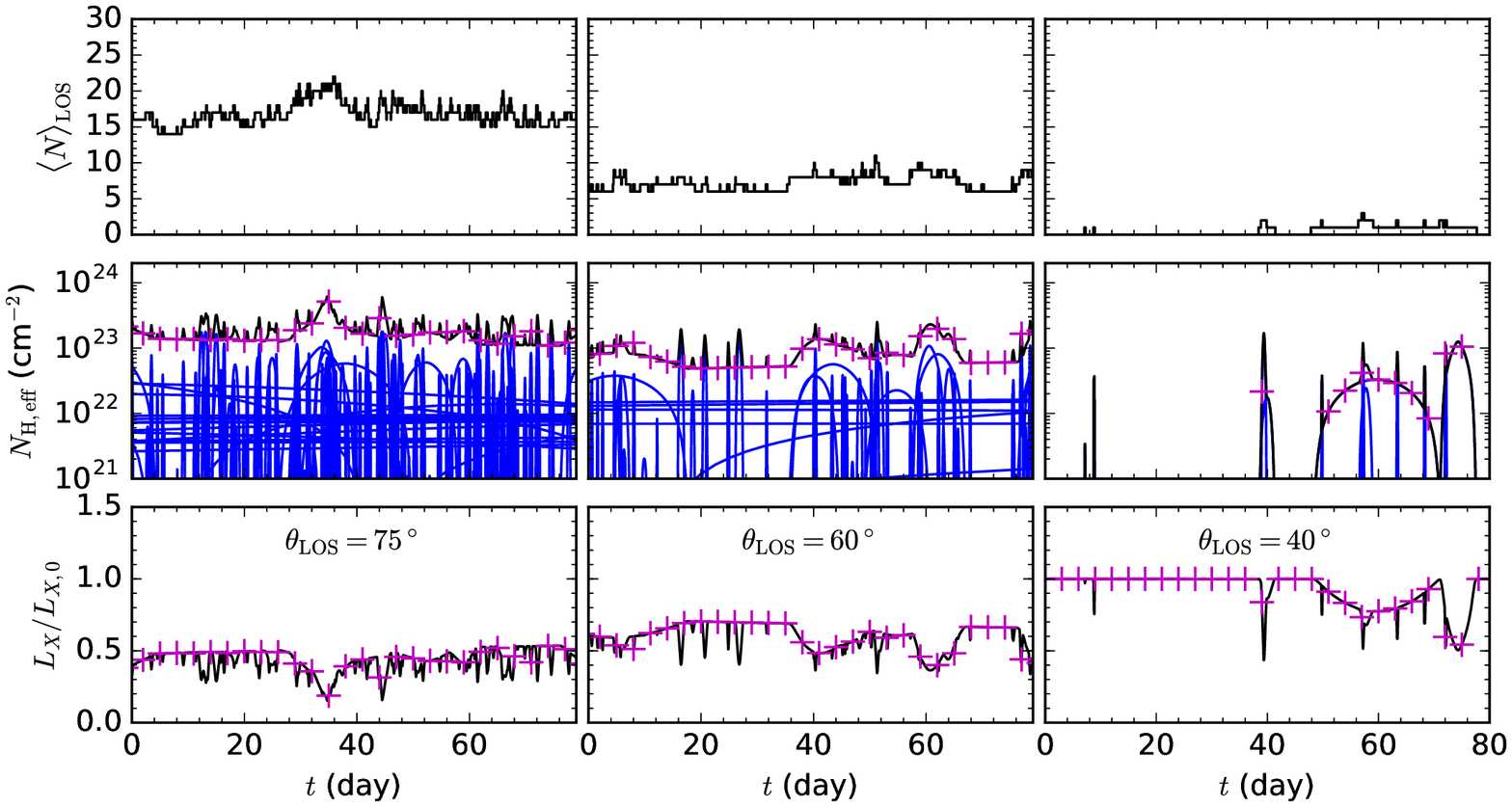}
\caption{Evolution curves for the number of the eclipsing clouds
($\langle N\rangle_\LOS$; top panels), the absorption column density
due to the eclipses ($N_{\rm H,eff}$; Eq.~\ref{eq:RAnu}; middle panels), and
the mock
observations of the $2-10$\,keV flux ($L_X$) relative to the intrinsic
$2-10$\,keV X-ray flux ($L_{X,0}$; bottom panels) obtained for 
model C1.  The black and the blue lines in the middle
panels represent the total column density and the contribution by
the individual clouds across the LOS, respectively.
The viewing angle $\theta_{\rm LOS}$ is set to $75\arcdeg$, $60\arcdeg$,
and $40\arcdeg$ in the left, middle, and right panels, respectively.
The magenta crosses in the middle and bottom panels
mimic the observational samples collected in every $3$ days.
As seen from the figure, the absorption at a given time is caused by multiple
clouds if $\theta_{\rm LOS}$ is high and mainly caused by a single cloud if
$\theta_{\rm LOS}$ is low.
}
\label{fig:f6}
\end{figure*}

The modeled X-ray light curves of an AGN is generated by a combination of
the transmitted flux of a power-law intrinsic X-ray emission through
the absorption cloud(s) and that of its reflection component. The
software XSPEC, version 12.8.1, is adopted to generate the mock spectra
for those modeled AGNs by $\rm
phabs*pexrav$\footnote{http://cxc.cfa.harvard.edu/ciao/}, where the
pexrav is the disk reflection model \citep{Magdziarz95} with a canonical
power-law index of $-1.8$, and the input column density is given by
$N_{\rm H,eff}(t)$ obtained from our simulations.
The X-ray flux variations $L_{\rm X}(t)$ at the $2-10$\,keV band can
then be obtained by integrating the mock X-ray spectra.  We find that
a slight deviation to the canonical power-law index leads to little
change in the model results. We also find negligible difference 
when changing the working model to $\rm phabs*powerlaw$, i.e., an absorbed pure 
power-law model, as the reflection component is $\lesssim 5\%$ of the total 
flux in the $2-10$\,keV band. For simplicity, the reflection component from individual
clouds is ignored. 

In this paper, we focus on the $2-10$\,keV band because the cloud
absorption of the X-ray emission in this band is sensitive to the
column density when $10^{21} \la N_{\rm H,eff} \la10^{24} \cm^{-2}$,
similar to the range of the effective absorption column density 
currently detected for the eclipsing AGNs \citep[e.g.,][]{Risaliti02}.
To demonstrate this, the left panel of Figure~\ref{fig:f4} shows the
mock observational spectra resulted from the several cases with column
densities of $10^{21}$, $10^{22}$, $10^{23}$, and $10^{24}{\rm
cm}^{-2}$, respectively; the right panel of Figure~\ref{fig:f4} shows
the ratio of the mock flux to the intrinsic X-ray flux in the
$2-10$\,keV band.  As seen from Figure~\ref{fig:f4}, if $N_{\rm H,eff}
\lesssim 2 \times10^{20} {\rm cm}^{-2}$, the absorption is
insignificant; and if $N_{\rm H,eff}\gtrsim4\times10^{24} {\rm
cm}^{-2}$, the source is Compton-thick and almost all X-ray radiation
at $2-10$\,keV are absorbed. The absorption column densities resulting
from all of the models discussed in this paper are indeed in the range
from $\sim 10^{21}$ to $10^{24}~{\rm cm}^{-2}$ in most of the cases.  

We obtain the variation curves for the mock X-ray flux $L_{\rm X}(t)$ and
the mock column density $N_{\rm H, eff}(t)$ with time intervals of
$\delta t$ within a period of $T_\tot$. Note that the PSDs derived
from the variations of the mock X-ray flux and absorption column
density (see Section~\ref{sec:pds}) are valid within the frequency range
$T_\tot^{-1}\la \nu\la \delta t^{-1}$.
We set $\delta t=10^{-3}$\,days and $T_\tot=10^{5} $\,days,
which covers the frequency range of the estimated PSDs for some AGNs,
with $\nu\sim 10^{-3}$--$10^{-9}$\,Hz,
\citep[e.g.][]{Mchardy87, Uttley02, Markowitz03, GonzlezMart12}. Note
that the adopted $T_{\rm tot}$ is longer than 
the currently available observation period on the long-term X-ray variations
of AGNs. Choosing a sufficiently long $T_{\rm tot}$ is reasonable for the
purpose of this paper so that the shape of the PSDs revealed from
our Monte-Carlo simulations in Section~\ref{sec:pds} below spans a sufficiently wide frequency
range.
With the adopted values of $\delta t$ and $T_\tot$, we have $\delta t \ll
T_{\rm B} \ll T_{\tot}$ ($T_{\rm B}=1/\nu_{\rm B}$ in Eq.~\ref{eq:bkpow}),
and the adopted time resolution is sufficiently
high and the total time duration is sufficiently long for the convergence of
the model results. 

\subsection{Results}
\label{subsec:results}

Figure~\ref{fig:f5} shows the statistical properties of the X-ray
eclipses obtained for those models listed in Table~\ref{tab:t1}. The curves
in the figure represent the results obtained from the analytical estimates
presented in Section~\ref{sec:analytic}. As the Monte-Carlo simulation results
are quite consistent with the corresponding analytical estimates, 
we only show the Monte-Carlo simulation results of one model (C1) in
Figure~\ref{fig:f5} for view clarity.  As seen from 
Figure~\ref{fig:f5}(a), the event rate $\mathscr{R}$ depends strongly on the
viewing angle $\theta_\LOS$ simply because of the dependence on the
function $\mathscr{H}(\theta_\LOS)$ (see Eq.~\ref{eq:ER}).  For those models
with the same viewing angle, $\mathscr{R}$ also strongly depends on the
average semimajor axis of the cloud ensemble $\langle a_{\rm
E}\rangle$ (Eq.~\ref{eq:aE}), which is determined by the radial and the size
distributions of the eclipsing clouds.
Choosing a different distribution of
the sizes of the eclipsing clouds and/or a different inner boundary
for the clouds ($a_{\rm c,min}$) leads to a different event rate. 
Those models with more clouds
located at the inner region (models C1-C4) generally have larger
event rates compared with those models with more clouds located at
the outer region (models B1 and B2). For models A1 and A2, the
majority of the clouds are located at intermediate distances to the
central MBH so that the event rates of the eclipses are also intermediate
compared with the other models.

Figure~\ref{fig:f5}(b) shows the mean number of the eclipsing
clouds ($\langle N \rangle_\LOS$) for different models, which are
compatible with the observational constraints for some AGNs
\citep[e.g.,][]{Nenkova08,Ichikawa15}.  As shown in this panel,
$\langle N \rangle_\LOS$ depends on the viewing angle because of the
assumed flattening of the spatial distribution for the parent
population of the eclipsing clouds. According to
Equation~(\ref{eq:nlos}), $\langle N \rangle_\LOS$ also depends on
$\langle a_N \rangle$, which is determined by $f_{a_{\rm c}}(a_{\rm c})$,
$R_{\rm c}(a_{\rm c})$, and $R_X$.  Note that the $\langle N \rangle_\LOS$
resulting from models A1, A2, B1, B2, and C1-C3 are similar for a
fixed viewing angle, as shown in panel (b). In all of these models,
$\alpha_{a_{\rm c}}=1$ and $\langle a_{\rm N} \rangle^{-2} \propto \int
[1+R_{\rm c}(a_{\rm c})/R_{\rm X}]^2 f_{a_{\rm c}}(a_{\rm c}) a^{-2}_{\rm c}
da_{\rm c} \sim $constant,  which coincidentally leads to quite similar
$\langle N \rangle_\LOS$ because $\langle N\rangle_\LOS \propto
\langle a_N\rangle^{-2}$. In model C4, $\alpha_{k_{\rm c}}=0.5$, i.e.,
the sizes of the majority of the clouds are set to be substantially
smaller than those in other models, and thus $\langle
N\rangle_\LOS $ is much smaller than that from other models. 

Figure~\ref{fig:f5}(c) and (d) show the distribution of the
time periods of eclipses [$f_{t_{\rm E}}(t_{\rm E})$; Eq.~\ref{eq:pdte}] 
and the distribution of the
semimajor axes of eclipsing clouds [$f_{a_{\rm c},\rm E}(a_{\rm c})$; Eq.~\ref{eq:facE}],
respectively. For models B1 and B2, relatively more clouds are located
at large distances, and thus there are relatively more eclipses with
long periods, compared with those of the other models.  For models
B2, C2, C3, and C4, the inner boundary for the clouds is smaller, and
thus there are relatively more eclipses with short periods ($\la
1$\,day), compared with those of the other models (A1, A2, B1, and
C1). The sharp decrease of $f_{t_{\rm E}}(t_{\rm E})$ at the short-timescale
end is mainly due to the cutoff of the semimajor axes at
$a_{\rm c,min}$ for the eclipsing clouds. As seen from
Figure~\ref{fig:f5},  $f_{t_{\rm E}}(t_{\rm E})$ and $f_{a_{\rm c},\rm E}(a_{\rm c})$ also
strongly depend on the size distribution of the clouds ($R_c(a_c)$).
For model C4, the sizes of clouds are set to be substantially
smaller than those clouds in other models, especially at large $a_{\rm
c}$, which leads to a sharp decrease of the number of eclipses with
large $a_{\rm c}$.

Figure~\ref{fig:f6} shows the evolution curves for the number of
eclipsing clouds ($\langle N\rangle_\LOS$; top panels),  the absorption column density ($N_{\rm H,eff}$; Eq.~\ref{eq:RAnu}; middle
panels), and the ratio of the ``observed'' X-ray flux to the intrinsic
X-ray flux ($L_X/L_{X,0}$; bottom panels) obtained from the simulations for model
C1, respectively.  The absorption column density contributed by each
individual eclipse cloud (the blue solid lines) are also shown in the
middle panels. Among them, the long-time variations of both the X-ray
flux and the column density are due to the eclipsing clouds in the
dusty torus; the short-time variations are due to the eclipsing clouds
in the BLR.  As seen from Figure~\ref{fig:f6}, the absorption
structures are complex if the viewing angle is large (e.g.,
$\theta_\LOS=75^\circ$) because the absorption at any given time can
be due to multiple clouds (the left middle and bottom panels), while
it is relatively simple if the viewing angle is small (e.g.,
$\theta_\LOS=40^\circ$) because the absorption is caused by individual
eclipsing clouds occasionally (the right middle and bottom panels).
According to Figure~\ref{fig:f6}, the presumption adopted in many
previous studies that the observed variations of the absorption column
density is due to {\it single} eclipsing clouds may not be realistic
\citep[e.g.,][]{Risaliti09, NR11, Sanfrutos13, Markowitz14}.

\subsection{Reconstruction and Decomposition of Individual Eclipses}
\label{subsec:bias}

In principle, the individual eclipsing events can be reconstructed by
using the evolution curves of the X-ray flux and the absorption column
density.  The information on the properties of those eclipsing clouds
may thus provide strong constraints on the spatial distribution and
the intrinsic properties of their parent population as demonstrated in
some previous studies \citep[e.g.,][]{Sanfrutos13}. However, it is
usually assumed in those studies that the variations of the absorption
column density and the X-ray flux are simply due to one single
eclipsing cloud. This simplified assumption may be violated in some
cases. For example, (1) the periods with and without absorption by
eclipses are mixed together when the time resolution of the X-ray
observations is not sufficiently high (e.g., substantially longer than
$\langle t_{\rm E} \rangle$), and thus the variation of the absorption
column density estimated from the observations is biased; (2) the
absorption may be complicated when more than one cloud crosses the LOS
at the same time (e.g., see the case with $\theta_\LOS=60\arcdeg$ for
model C1 in Fig.~\ref{fig:f6}), and the assumption of one single
cloud eclipsing may lead to an underestimation of the total number of the
clouds and significant biases in the estimation of the intrinsic
properties of those clouds.

Whether the properties of individual eclipsing clouds can be
accurately reconstructed from the variation curves of the X-ray
flux/luminosity depends on the mean number of the eclipsing clouds at
a given moment even if the time resolution is sufficiently high. The
properties of the eclipsing clouds may be securely reconstructed from
the X-ray flux curve if $\langle N \rangle_\LOS \la 1-2$. However,
they are difficult to constrain if $\langle N \rangle_\LOS >2 $
because of the complicated and irregular absorption due to multiple
eclipsing clouds at almost any given moment. As shown in
Figure~\ref{fig:f6}, $\left<N\right>_\LOS$ can be larger than 2 for
some settings on the spatial distribution and the intrinsic properties
of the parent population of the eclipsing clouds.  Therefore, it is
not easy to directly extract the properties of the eclipsing clouds from
the variation curves of the absorption column density and the X-ray
flux.

\section{Power Spectral Density of the X-ray Flux Variation}
\label{sec:pds}

\begin{figure}
\centering
\includegraphics[scale=0.8]{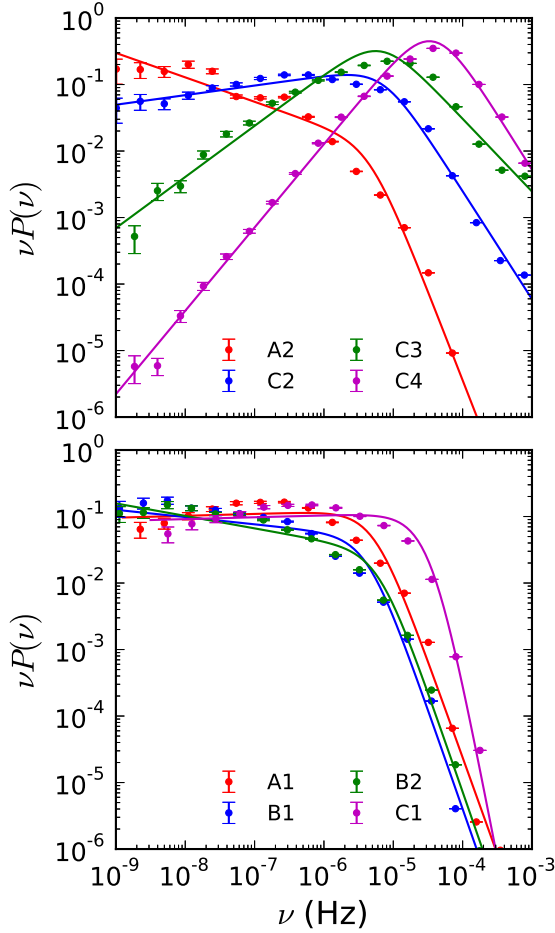}
\caption{Smoothed PSDs ($P(\nu)$) resulting from the mock
observations of the $2-10$\,keV X-ray flux variations for different
models. The top panel shows the results obtained for models A2 (red), C2 (blue),
C3 (green), and C4 (magenta),
respectively; the bottom panel shows the results obtained for models
A1 (red), B1 (blue), B2 (green), and C1 (magenta), respectively.
The error bars associated with each data point
show the Poisson errors due to the limited number of points in each
logarithmic frequency bin. The solid lines with different colors
represent the best fit of a double power law to the points of each model (see
Eq.~\ref{eq:bkpow}).
The model results are independent of the viewing
angle $\theta_\LOS$. 
The PSD indicates a breaking power-law distribution.
See Section~\ref{sec:pds}. 
}
\label{fig:f7}
\end{figure}

The PSD is a powerful tool to investigate the nature of the flux
variation at a given band for AGNs and X-ray binaries
\citep[e.g.,][]{Uttley02, Markowitz03, GonzlezMart12, Cui97}. In this
section, we investigate the dependence of the PSD extracted from the
mock variations of $L_{\rm X}$ and $N_{\rm H,eff}$ on the model
parameters listed in Table~\ref{tab:t1}. For a given variation curve
of the `observed' $2-10$\,keV X-ray flux  or $N_{\rm H,eff}$, we first
obtain its discrete Fourier transform, $F(\nu)$.  Given the total
observational duration $T_\tot$ and the sampling interval $\delta t$,
then the total number of the data points is $N=T_\tot/ \delta t$.
The frequencies have $N/2$ discrete values,
i.e., $\nu_j=j/T_\tot$ with $j=1,2,... N/2$.  The normalized
PSD of the variation curve can then be obtained by
\be
P(\nu_j)=\frac{2T_\tot}{\sigma^2 N^2}\left| F(\nu_j) \right|^2,\quad (j=1,...,N/2),
\label{eq:pds}
\ee
where $\sigma$ is the variance of the curve with substantially large
$T_{\rm tot}$ and small $\delta t$, i.e., $\sigma^2 =\displaystyle{
\frac{1}{N} \sum^N_{i=1}} (L_{{\rm X}}(t_i)-\langle L_{\rm
X}\rangle)^2$, $t_i$ represents the time at the $i$th time interval,
and $\langle L_{\rm X}\rangle=\displaystyle{\frac{1}{N}\sum^{N}_{i=1} } L_{{\rm X}}(t_i)$ is the mean.
Note that the normalized PSD defined above is slightly different from
those PSDs adopted in the literature (denoted by $P'(\nu)$), with $P'(\nu) =
\frac{2T_\tot}{\mu^2 N^2} \left|F(\nu)\right|^2 =
\frac{\sigma^2}{\mu^2} P(\nu)$, where $\mu =\langle L_{\rm X}\rangle$
\citep[e.g.,][]{Markowitz03, Uttley02}.  With the
definition in Equation~(\ref{eq:pds}), the PSDs resulting from
different variation curves all satisfy $\int P(\nu) d\nu =1$, which
enables us to focus on the PSD shape without involving other
complexities caused by the variations as discussed later
in Section~\ref{subsec:var}. 

The $P(\nu)$ directly estimated from Equation~(\ref{eq:pds}) contains
a total number of $N/2\sim10^7$ points and is usually noisy. To reduce
the noise, we divide the logarithm of the frequency $\nu$ into $20$ bins with an equal
logarithmic interval $\delta\log \nu$, and in each bin of $\log \nu_k-\delta \log \nu/2
\rightarrow \log \nu_k+\delta\log \nu/2$ ($k=1,2,...20$), we set $P(\log \nu_k)$ to be the
mean value within this frequency bin.  In Section~\ref{subsec:shape}, we fit
the shape of the PSD mainly in the range of $\nu$ from $3/T_{\tot}$ to
$1/(6\delta t)$.
We find that the PSDs resulting from those models listed in
Table~\ref{tab:t1} can be approximately described by a double
power law, typically within the frequency range from $\sim 10^{-3}$ to
$10^{-9}$\,Hz, as shown in Figure~\ref{fig:f7}. At the low-frequency
end, i.e., $\nu \ll 10^{-9}$\,Hz, the PSDs drop sharply, mainly due to
the cutoff of the distribution of clouds at an outer boundary $a_{\rm
c,max}$; at the high-frequency end, i.e., $\nu \ga 10^{-3}$\,Hz, the
PSDs may fluctuate if the time interval of the measurements $\delta t$
is not sufficiently small. We do not include the sharply dropping
part of the PSDs at low frequencies and the fluctuating part at high
frequencies when using a double power-law form to fit the PSDs below.

\subsection{The Shape of the Power Spectrum Density}\label{subsec:shape}

In the absorption scenario presented above, the shape of the PSD
contains the information on the spatial distribution and the intrinsic
properties of the eclipsing clouds and their parent population.
We choose the
following double power-law form to describe the PSD shapes of the
X-ray flux variations resulting from the models listed in
Table~\ref{tab:t1}, i.e., 
\be
P(\nu)=A\left(\frac{\nu}{\nu_{\rm B}}\right)^{\gamma_{\rm l}}\left[1+
\left(\frac{\nu}{\nu_{\rm B}}\right)^{2} \right]^{(\gamma_{\rm
h}-\gamma_{\rm l})/2},
\label{eq:bkpow}
\ee
where $A$ is the normalization of the PSD, $\gamma_{\rm l}$ and
$\gamma_{\rm h}$ are the slopes of the PSD at the low frequencies and the high
frequencies, respectively, and $\nu_{\rm B}$ is the break frequency at
the turning point. The break timescale, corresponding to the
break frequency, is defined as $T^{L_{\rm X}}_{\rm B}\equiv1/\nu_{\rm
B}$. Similarly, the PSDs for the absorption column density $N_{\rm
H,eff}$ variations can also be fitted by a double power-law function,
the same as that shown by Equation~(\ref{eq:bkpow}); and for these
cases we denote the slopes at the low frequencies and the high frequencies by
$\beta_{\rm l}$ and $\beta_{\rm h}$, respectively, and the break
timescale by $T_{\rm B}^{N_{\rm H}}\equiv 1/\nu_{\rm B}$.  

The best-fit values of $\gamma_{\rm l}$, $\gamma_{\rm h}$, and
$T^{L_{\rm X}}_{\rm B}$ (or $\beta_{\rm l}$, $\beta_{\rm h}$, and
$T^{N_{\rm H}}_{\rm B}$) for those PSDs resulting from the mock
observations for different models are listed in Table~\ref{tab:t1}. 
We find that the PSD shapes are independent of $\langle N \rangle_\LOS$,
if the absorption column density is in the range of $10^{21} - 10^{24}
\cm^{-2}$. Figure~\ref{fig:f7} shows the PSDs obtained from the mock
observations for models A2, C2, C3, and C4 (top panel), and models
A1, B1, B2, and C1 (bottom panel), respectively.  Our calculations show
that the PSD shapes resulting from the mock observations on the X-ray
flux and those from the absorption column density are quite similar
(see the best-fit values for the PSD shapes in Table~\ref{tab:t1}). We
find that such a similarity remains if the effective absorption column
density is in the range of $10^{21} - 10^{24} \cm^{-2}$,  in which the
mock X-ray flux is sensitive to the variation of $N_{\rm H,eff}$.
Note that the sizes and the hydrogen densities of those eclipsing clouds with the
same semimajor axis $a_{\rm c}$ are probably not exactly the same,
which is not considered in our calculations.  If considering the
scatters of those model parameters, then the resulting PSDs near the
break frequency should be smoothed slightly, and therefore the
break frequency and the power-law slopes at both the high and the low-frequency
ends may also change slightly, compared with those obtained without
considering the scatters of the sizes and hydrogen densities.

A number of studies based on the PSD analysis of the X-ray variations
suggested that the PSDs of AGNs have a slope of $\gamma_{\rm l} \sim -1$
at the low frequencies and $-4<\gamma_{\rm h} \leq-2$ at the high frequencies
\citep[e.g.,][] {Markowitz03, Uttley02, GonzlezMart12}.  According to
Table~\ref{tab:t1}, models A1, B1, B2, C1, and C2 can produce PSDs
similar to the observational ones. 
Those models with less column density variations caused
by the clumps in torus, e.g., model C3 (or C4), result in
$\gamma_{\rm l}=-0.2$ (or $0.3$ ), inconsistent with observations.  
Since the X-ray variations in some
AGNs, e.g., NGC 1365 \citep[e.g.][]{Risaliti99}, NGC 7582
\citep[e.g.][]{Risaliti02}, and NGC 4151 \citep{Schurch02,
Markowitz14}, are probably dominated by the absorption of eclipsing
clouds, therefore, the observationally determined $\gamma_{\rm l} \sim
-1$ at the low frequencies (corresponding to long timescales) for the AGN
PSDs may support the scenario in which the eclipsing due to the clumps in the
dusty torus plays an important role in the column density and flux
variations.
In principle, the observational measurements on the PSDs of AGNs can put
strong constraints on the spatial distribution and intrinsic
properties of the eclipsing clouds and their parent population, which
may be obtained by future X-ray observations and missions.

\begin{figure*}
\centering
\includegraphics[scale=0.8]{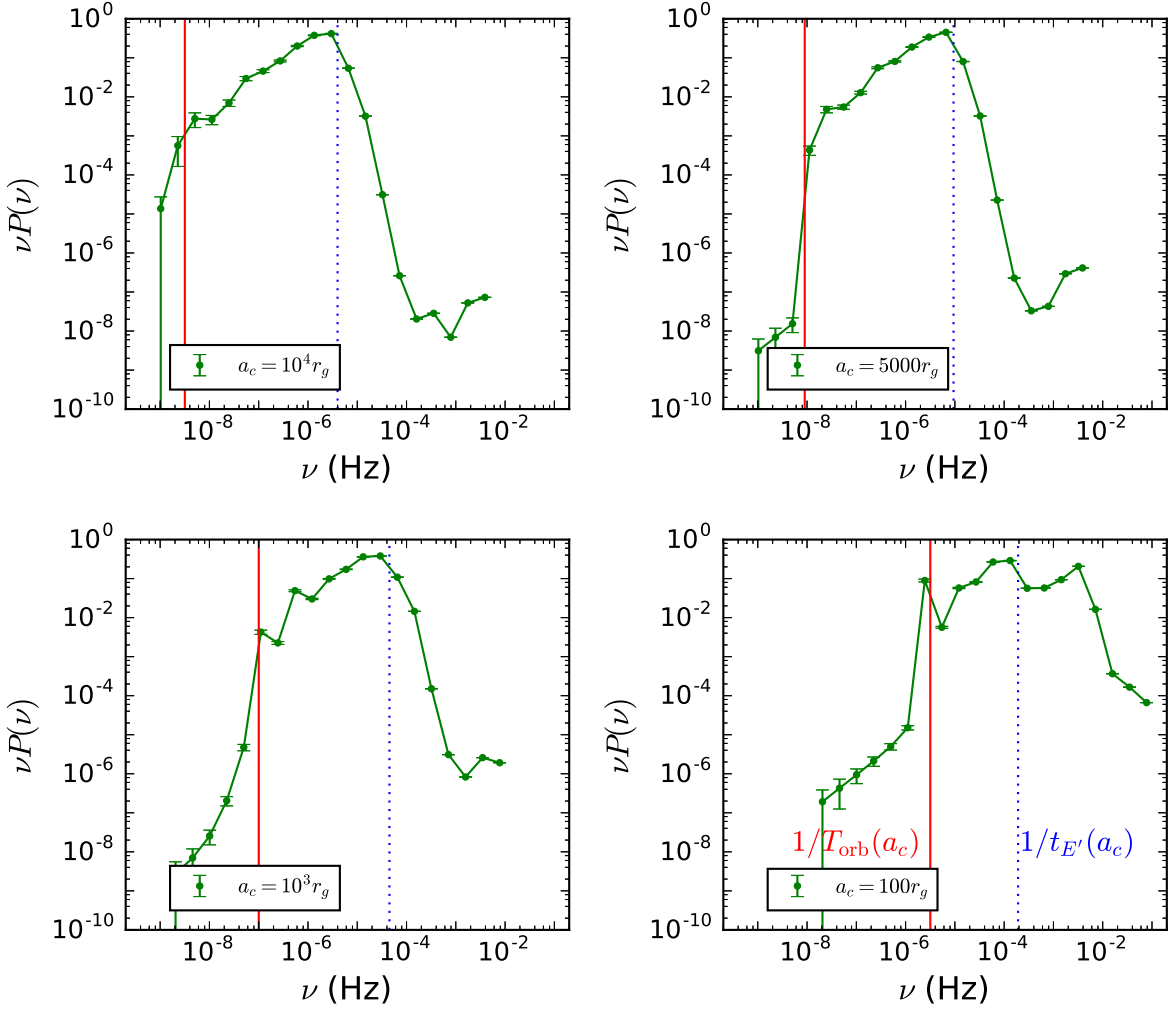}
\caption{The PSDs [$P(\nu)$] resulting from a model in which
all the clouds are set to have the same semimajor axis and all the other
model parameters are similar to those in model C2.
The semimajor axes of the clouds are $10^4r_\g$ (top-left
panel), $5\times 10^3 r_\g$ (top-right panel), $10^3r_\g$ (bottom-left
panel), or $10^2r_\g$ (bottom-right panel), as labeled in each panel.
The error bars associated with each data point show the Poisson errors
due to the limited number of points in each logarithmic frequency bin.
The vertical red solid line and the blue dotted line in each panel
mark the positions of $\nu=1/T_{\rm orb}(a_{\rm c})$ and
$\nu=1/t_{\rm E'}(a_{\rm c})$, respectively, where $T_{\rm orb}(a_{\rm c})$ is
the orbital period of a cloud with semimajor axis of $a_{\rm c}$,
and $t_{\rm E'}(a_{\rm c})$ is the maximum eclipsing timescale given by
Equation~(\ref{eq:te}). 
This figure illustrates that the break shown in the PSD $P(\nu)$ is dominated
by the characteristic timescales $t_{\rm E'}(a_{\rm
c})$ and $T_{\rm orb}(a_{\rm c})$.
See Section~\ref{subsec:tb}.
}
\label{fig:f8}
\end{figure*}

The X-ray flux variation resulting from any of the models listed in
Table~\ref{tab:t1} is dominated by those eclipsing events with $b\sim
0$, which lead to relatively larger column density variations. If the
X-ray variation is mainly contributed by eclipsing clouds with the
same $a_{\rm c}$, then the corresponding PSD approximately peaks at a
frequency $\sim\nu'_{\rm E}(a_{\rm c})\equiv 1/t'_{\rm E}(a_{\rm c})$. 
Here $t'_{\rm E} (a_{\rm c})$
is the maximum eclipsing time of the clouds with semimajor axes
$a_{\rm c}$,  which is given by Equation (\ref{eq:te}).
Figure~\ref{fig:f8} shows the PSD derived from the mock X-ray flux
variations due to eclipses by clouds with the same semimajor axes,
i.e., $a_{\rm c}=10^2$, $10^3$, $5\times 10^3$, and
$10^4\, r_\g$, respectively. As seen from this figure, our numerical
results demonstrate that the PSDs do peak at $\nu\sim 1/t'_{\rm
E}(a_{\rm c})$.

If the PSD of the X-ray variation is contributed by eclipsing clouds
with various $a_{\rm c}$, the magnitude of the PSD at a given
frequency $\nu$ then approximately corresponds to the X-ray
variation contributed from those eclipsing clouds at $a_{\rm c}=t_{\rm E'}^{-1}(1/\nu)$ (here $t_{\rm E'}^{-1}$ is the inverse function of
$t_{\rm E'}$).  Therefore, the PSD at low frequencies is determined by
the eclipsing clouds at the outer region, while at high frequencies it
is determined by the clouds at the inner region. The detailed
dependence of the shape of the PSD (i.e., a double power-law shape
described by $T_{\rm B}$, $\gamma_{\rm l}$, and $\gamma_{\rm h}$) on
the model parameters is discussed as follows.

\subsubsection{The Break Timescale $T_{\rm B}$}
\label{subsec:tb}

Table~\ref{tab:t1} lists the break timescale
corresponding to the break frequency $\nu_{\rm B}$ of the PSDs for
the mock X-ray variations ($T_{\rm B}^{L_X}$) and also those for the mock column density
variations ($T_{\rm B}^{N_{\rm H}}$) resulting from different models.
Assuming that the mass of the central MBH is $10^7\msun$, the break
timescale $T_{\rm B}^{L_{\rm X}}$  obtained from different models range
from $0.3$ to $3$\,days,  which appears roughly consistent with the
observations \citep{Markowitz03, McHardy06, GonzlezMart12}.

Since the PSD at the low (or high) frequencies is mainly determined by the
clouds at the outer (or inner) region, the shape of the PSD due to
absorption may be bent over at several frequencies due to the
existence of the following characteristic scales.
\begin{itemize}
\item The break radius $r_{\rm t}$ in the spatial distribution of
the clouds (Eq.~\ref{eq:fa}): if $\alpha_{a_{\rm c},1} \ne \alpha_{a_{\rm c},2}$, the
radial distribution of those eclipsing clouds in the region inside
$r_{\rm t}$ is different from that in the region outside $r_{\rm t}$.
\item The inner boundary for the spatial distribution of the eclipsing
clouds, i.e., $a_{\rm c,min}>0$: no eclipsing clouds exist within
$a_{\rm c,min}$, and therefore, the PSD  drops rapidly at frequencies
$\nu \ga 1/t_{\rm E'}(a_{\rm c,min})$.
\item The characteristic semimajor axis $a_{\rm eq}$ at which the
cloud size $R_{\rm c}$ is equal to the X-ray source size
$R_{\rm X}$:  for an eclipsing cloud with $a_{\rm c} < a_{\rm eq}$ and
$R_{\rm c}\lesssim R_{\rm X}$, the resulting absorption column density
approximately follows $N_{\rm H}\propto R_{\rm c}^3 /R_{\rm X}^2$;
while for an eclipsing cloud with $a_{\rm c} \gg a_{\rm eq}$ and
$R_{\rm c}(a_{\rm c})\gg R_c(a_{\rm eq})$, the column density approximately
follows $N_{\rm H} \propto R_{\rm c}$. The dependence of $N_{\rm H}$
on $R_{\rm c}$ changes smoothly around $a_{\rm eq}$. According to our
numerical results, we find that the turning points of the PSD are
around $\nu\sim 1/ t_{\rm E'}(a_{\rm eq})$, where $a_{\rm eq}=R_{\rm
c}^{-1}(\xi  R_{\rm X}) =a_{\rm c,min}(\xi R_{\rm
X}/R_{\rm c,0})^{1/\alpha_{R_{\rm c}}}$, and 
$\xi$ is slightly model-dependent. For model C1, we have
$\xi\simeq1.7$. For other models listed in Table~\ref{tab:t1}, we have
$\xi\sim0.3-2.0$.
If $R_{\rm c,0}<\xi R_{\rm X}$,  we have $a_{\rm eq} \ge a_{\rm c, min}$.
However, $a_{\rm eq}$ does not exist if $R_{\rm c,0}>\xi R_{\rm X}$.
\end{itemize}

We find that the resulting PSDs from those models listed in
Table~\ref{tab:t1} bend over slightly near $\nu_{\rm E'}(r_{\rm t}) =
1/t_{\rm E'}(r_{\rm t})$ if $\alpha_{a_{\rm c},1}\ne \alpha_{a_{\rm c},2}$.  For
those models listed in Table~\ref{tab:t1}, we have $\nu_{\rm E'}(r_{\rm t})
\ll \nu_{\rm B}$. Our model results show that this bending of the PSDs
around $\nu_{\rm E'}(r_{\rm t})$ is not significant compared with that
near $\nu_{\rm B} = 1/T_{\rm B}$. The main reason is that the
difference between $\alpha_{a_{\rm c},1}$ and $\alpha_{a_{\rm c},2}$ is only
moderate.  However, the radial distributions of eclipsing clouds for
models with extremely large differences between $\alpha_{a_{\rm c},1}$ and
$\alpha_{a_{\rm c},2}$ are usually unphysical. For example, the cases with
clouds extremely abundant at the outer region of the dusty torus while
no clouds are in the BLR, or vice versa, are inconsistent with
observations and the unification model for AGNs.

\begin{figure*}
\centering
\includegraphics[scale=0.6]{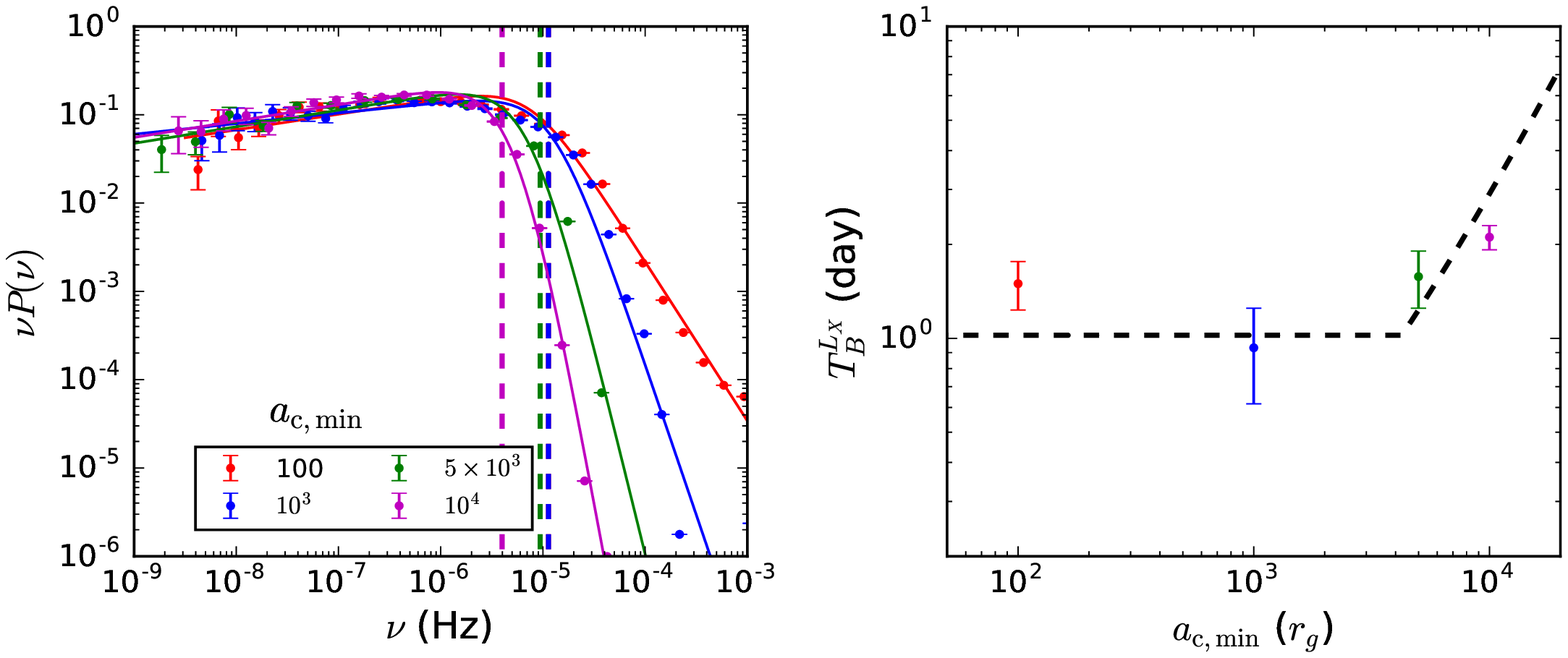}
\includegraphics[scale=0.6]{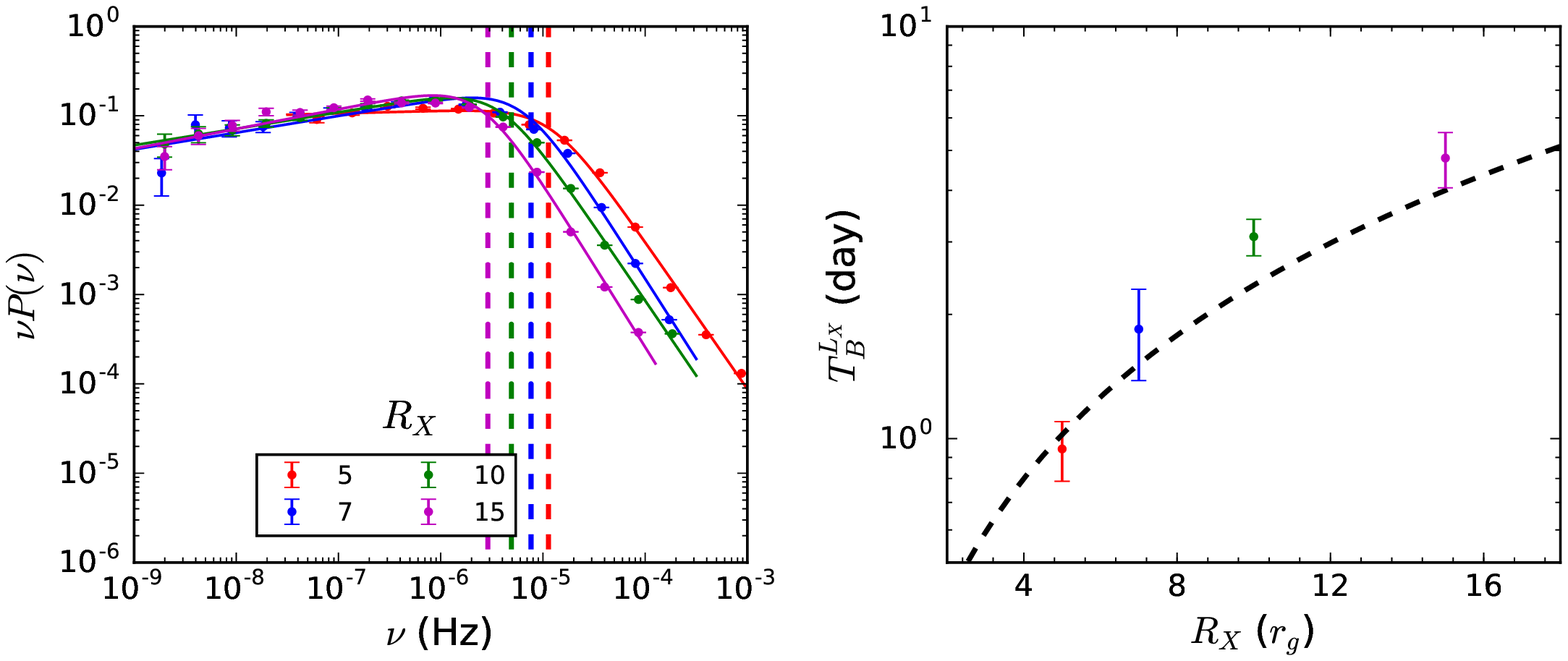}
\caption{Top-left panel: the PSDs resulting from model C2 and
three other models with parameters similar to those in model C2 except for a
different value of the inner boundary of the spatial distribution of
those clouds as labeled in the figure  ($a_{\rm c,min} =10^3$, $5
\times 10^3$, and $10^4r_\g$, respectively). The vertical red dashed line
overlaps with the vertical blue dashed line in this panel. 
Top-right panel: the
break timescales obtained from the fittings to the PSDs shown in the top-left
panel.  Bottom-left panel: the PSDs resulting from model C2
and three other models with parameters similar to those in model C2 except for
a different value for the size of the X-ray source as labeled in the
figure ($R_{\rm X}=7$, $10$, and $15r_\g$, respectively).  Bottom-right
panel: the break timescales obtained from the fittings to the
PSDs shown in the bottom-left panel. The solid lines in the left
panels represent the best fits to the PSDs.  The vertical dashed lines
in the left panels represent the estimates obtained from
Equation~(\ref{eq:nub}). The black dashed curves in the right panels
represent the estimates obtained from Equation (\ref{eq:nub}). 
The simulation results are consistent with Equation (\ref{eq:nub}) well, which
suggests that the break frequency in the PSD is mainly affected by the clouds with
semimajor axis $a_c\sim \max(a_{c,\min},a_{\rm eq})$.
See Section~\ref{subsec:tb}.
}
%
\label{fig:f9}
\end{figure*}

We find that the resulting PSDs break significantly near $\nu_{\rm
E'}(a_{\rm c,min})=1/t_{\rm E'}(a_{\rm c,min})$ due to the existence
of an inner boundary ($a_{\rm c,min}$) for the cloud spatial
distribution, or near $\nu_{\rm E'}(a_{\rm eq})=1/t_{\rm E'}(a_{\rm
eq})$ due to the existence of the characteristic scale $a_{\rm eq}$.
If both $a_{\rm c,min}$ and $a_{\rm eq}$ exist, then the break
frequency $\nu_{\rm B}$ obtained from the fitting to the PSDs is
roughly given by the one with a smaller value, i.e., 
\begin{equation}
\nu_{\rm B} \sim \min[ \nu_{\rm E'}(a_{\rm c,min}), \nu_{\rm
E'}(a_{\rm eq})],
\label{eq:nub}
\end{equation}
which appears as the first significant break in the PSDs with
$\nu>10^{-8}$ Hz, while the higher frequency one indicates a further 
decrease of the PSD magnitude at frequencies $\nu>\nu_{\rm B}$ (the
slope of those PSDs at higher frequencies becomes steeper). 
Figure~\ref{fig:f9} illustrates the dependence of the break
frequency $\nu_{\rm B}$ and the break timescale $ T_{\rm B}$ on $R_{\rm
X}$, $a_{\rm c,min}$, and $\alpha_{R_{\rm c}}$, as suggested by
Equation~(\ref{eq:nub}).  As seen from this figure, $T_{\rm B}$
increases with increasing the inner boundary of the radial
distribution of the eclipsing clouds ($a_{c, \rm min}$, top-left panel),
or with increasing the source sizes  ($R_{\rm X}$, bottom-left panel).
The right panels of  this figure show that $T_{\rm B}$ increases from
$\sim 1$\,day to $2$\,days when $a_{\rm c, min}$ increases from $100
r_\g$ to $10^4 r_\g$, and $T_{\rm B}$ increases from $\sim 1$ day up
to $\sim 4$ days when $R_{\rm X}$ increases from $5 r_\g$ to $15
r_\g$. These results are well consistent with the predictions by
Equation~(\ref{eq:nub}). 

The PSDs obtained from our models usually turn over smoothly around
the break frequency $\nu_{\rm B}$. However, the PSDs obtained from
observations are sometimes limited by the total observational duration
($T_{\rm tot}$) at the lowest frequency ($\nu_{\rm min} \sim 1/T_{\rm
tot}$) and the minimum sampling interval ($\delta t$) at the highest
frequency ($\nu_{\rm max} \sim 1/(2\delta t)$). If $\nu_{\rm min}$ (or
$\nu_{\rm max}$) is close to $\nu_{\rm B}$, then the $\gamma_{\rm l}$ (or
$\gamma_{\rm h}$) and $\nu_{\rm B}$ obtained from the best fits to the
PSDs are biased.  It is important to perform observations with a
sufficiently long period and a sufficiently small time interval in
order to estimate the shape of the PSD for the X-ray variation of an
AGN accurately. 

Note that the existence of the outer boundary for the spatial
distribution of those eclipsing clouds may also lead to a break in the
PSD. This break, however, corresponds to a timescale of $\gg 10$
years, which is probably difficult to probe in the near future.

\subsubsection{The PSD Shape at Low Frequencies: $\gamma_l$}
\label{subsec:gammal}

The slope of the PSD at frequencies $< \nu_{\rm B}$ is mainly
determined by the spatial distribution and the intrinsic properties of
the eclipsing clouds at the outer region, i.e., the dusty torus. If
the clouds at the outer region in a model have relatively larger sizes
(larger $\alpha_{R_{\rm c}}$) and hydrogen densities (larger $\alpha_{n_{\rm H}}$), or they are
more abundant, compared with those in another model, then the PSD
slope at low frequencies ($\gamma_{\rm l}$ or $\beta_{\rm l}$)
resulting from this model should be steeper (smaller) than those from
the other model. For example, $\gamma_{\rm l}$ resulting from the
model A2 is smaller than that from model A1 simply because the
hydrogen densities of the clouds in the dusty torus region in model A2 are
relatively larger compared with those in model A1. Similarly,
$\gamma_{\rm l}$ resulting from model C3 or C4 is larger than that
from model C2 because the hydrogen densities of the eclipsing clouds at
the outer region in model C3 are relatively smaller than those in
model C2 or the sizes of those clouds in model C4 are
relatively smaller than those in model C2. As seen from
Figure~\ref{fig:f7}, the magnitude of the PSD at lower frequencies
resulting from model B1 is slightly larger than that from 
model C1 (smaller $\gamma_{\rm l}$) because there are relatively more
clouds distributed at the outer region in model B1 than those in
model C1. Increasing the value of $\alpha_{R_{\rm c}}$, $\alpha_{n_{\rm H}}$,
$\alpha_{a_{\rm c},2}$, $\lambda_{\rm Edd}$, or $r_{\rm t}$, the contributions
from those eclipsing clouds at large distance to the central MBH to
the PSD increases correspondingly, which results in a smaller
$\gamma_{\rm l}$.  As an example, for model C2, $\gamma_{\rm l}$
decreases from  $\sim -0.8$ to $-1.2$ if $r_{\rm t}/r_{\rm d}$
increases from $0.01$ to $1$; while $\gamma_{\rm l}$ decreases from
$\sim 0.3$ to $\sim -1.3$ if $\alpha_{R_{\rm c}}$ increases from $0.5$ to
$1.2$; for model C1, $\gamma_{\rm l}$ decreases from $-0.2$ to
$-1.4$ if $\alpha_{n_{\rm H}}$ increases from $-2.5$ to $1.0$; while
$\gamma_{\rm l}$ decreases from $-0.8$ to $-1.2$ if $\lambda_{\rm Edd}$
increases from $0.01$ to $1$.  We find that the
dependence of $\gamma_{\rm l}$ on other parameters, i.e., $R_{\rm X}$, $a_{\rm
c,min}$, and $\alpha_{a_{\rm c},1}$, is little (see Fig.~\ref{fig:f9}).

Note also that the $\gamma_{\rm l}$ and $\gamma_{\rm h}$ of PSDs are
independent of $n_{\rm H,0}$ if the column density $N_{\rm H}$ ranges
from $10^{21}$ to $10^{24} \cm^{-2}$, as the increase of $n_{\rm H,0}$
enhances the contribution to the PSDs at both low and high frequencies
simultaneously.

\subsubsection{The PSD Shape at High Frequencies: $\gamma_{\rm h}$}
\label{subsec:gammas}

The magnitude of the resulting PSD at high frequencies, usually above
$10^{-5} -10^{-4}$\,Hz for an MBH with mass $10^7\msun$, is mainly
determined by the eclipsing clouds (e.g., the BLR clouds) at close
distances to the MBH. If the eclipsing clouds at the inner region in a
model have relatively larger sizes and densities, or if they are more
abundant at small $a_{\rm c}$, compared with those in another model,
then the magnitude of the resulting PSD ({\it not the slope}) at high
frequencies is larger compared with the other model. As discussed in
Section~\ref{subsec:tb}, the steep bending-down of the PSD at high
frequencies is mainly due to the lack of eclipsing clouds within
$a_{\rm c,min}$ and the less significant absorption due to those
clouds with smaller $a_{\rm c}$ because of their smaller size and
consequently smaller absorption column densities ($N_{\rm H}\propto
R_{\rm c}^3(a_{\rm c}) /R_{\rm X}^2$). The steep slope of the PSD at
high frequencies $\gamma_{\rm h}$ thus indicates the degree of the
lack of those eclipsing events with small distances to the MBH (or
small $a_{\rm c}$). We find that $\gamma_{\rm h}$ mainly depends on
$a_{\rm c,min}$ and the index $\alpha_{R_{\rm c}}$ describing the size
change of the eclipsing clouds with the semimajor axis. As shown in
Figure~\ref{fig:f9}, for a larger inner boundary ($a_{\rm c,min}$)
of the cloud distribution, the slope of the PSD at high
frequencies is steeper.
The $\alpha_{a_{\rm c},1}$ affects the value of $\gamma_{\rm h}$.
Decreasing $\alpha_{a_{\rm c},1}$ increases the magnitude of the PSD at
frequencies higher than the break frequency and lower than
$\nu_{\rm E'}(a_{\rm c,min})$ (e.g., see the PSDs of models A1
and C1 in the frequency range $\sim10^{-5}-10^{-3}$ Hz in Fig.~\ref{fig:f7}),
which may lead to a smaller $\gamma_h$ in the double power-law fit to the PSD
if $a_{\rm min}$ is only several times smaller than $a_{\rm eq}$ as in the
case for model C1.
As seen from Figure~\ref{fig:f7},
$\gamma_h$ in model C1 is steeper than that in model A1,
although $\alpha_{a_{\rm c},1}$ in
model C1 is relatively smaller, which is because no clouds are set within the
inner boundary $a_{\rm c,min}=10^3r_g$ and the decrease of $\alpha_{a_{\rm
c},1}$ leads to a larger change between the numbers of the clouds contributing
to the PSD at $\nu\sim10^{-5}-10^{-4}$ Hz and those contributing to the PSD
at higher frequencies.
If $a_{\rm min}\ll a_{\rm eq}$, decreasing $\alpha_{a_{\rm c},1}$ may increase 
the slope of $\gamma_h$ within the frequency range
$\nu_{\rm E'}(a_{\rm eq})\la\nu\ll\nu_{\rm E'}(a_{\rm c,min})$.
If $a_{\rm min}>a_{\rm eq}$, the dependence of $\gamma_h$ on $\alpha_{a_{\rm c},1}$ is weak.
The $\gamma_{\rm h}$ also depends on $\alpha_{n_{\rm H}}$ (models C2 and C3).
The $\gamma_{\rm h}$ is insensitive to $\lambda_{\rm Edd}$ and $r_t$,
as $\nu_{\rm E'}(r_{\rm t}) \ll \nu_{\rm B}$.
The value of $R_X$ may affect the position of $\nu_B$, but not $\gamma_{\rm h}$.

\subsection{The Relationship between $T_{\rm B}$ and $\bh$}
\label{subsec:relationship}
\begin{figure*}
\centering
\includegraphics[scale=0.85]{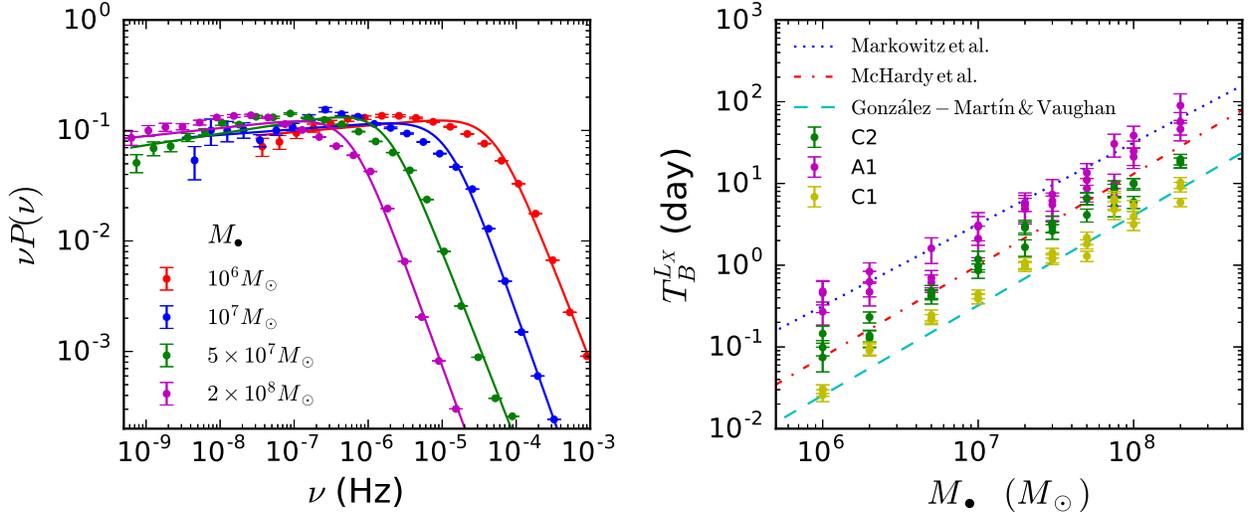}
\caption{Left: the PSDs resulting from model C2 by setting four
different values for the MBH mass and the best fits to them.
This panel shows that the shape of the PSD is insensitive to the MBH mass
except that the location of the break frequency decreases with
increasing MBH mass. 
Right: the break timescale $T_{\rm B}^{L_{\rm
X}}$  as a function of the MBH mass $\bh$.  The green, magenta, and
yellow circles represent the results obtained from models C2, A1,
and C1, respectively. The error bars associated with each point
represent the $1-\sigma$ error of the best fit. For each model, we
adopt various values for the mass of the central MBH in the range from
$10^6$ to $2\times10^8\msun$. For each given MBH mass, the
results obtained for those cases with three different view angles, i.e.,
$\theta_\LOS=75^\arcdeg, 60^\arcdeg$, and $45^\arcdeg$, are shown for
each model. The blue dotted line, the red dash-dotted line, and the
cyan dashed line represent the observational fitting results given by
\citet{Markowitz03}, \citet{McHardy06}, and \citet{GonzlezMart12},
respectively. 
This figure shows that the absorption scenario for
the AGN X-ray variations naturally leads to a strong correlation
between $T_{\rm B}^{L_{\rm X}}$ and $\bh$, which can be consistent
with the relationship obtained from observations.
See Section~\ref{subsec:relationship}.
}
\label{fig:f10}
\end{figure*}

A number of observations suggest that the break timescales $T_{\rm
B}^{L_{\rm X}}$ obtained from the PSDs of the AGN X-ray variations
correlate with the masses of the central MBHs. For example,
\citet{Markowitz03} found a linear correlation between these two
quantities as $T_{\rm B}^{L_{\rm X}}/{\rm day}=\bh/10^{6.5}\msun$;
\citet{McHardy06} confirmed the correlation between $T_{\rm B}$ and
$\bh$, and they further suggested that the accretion rate should also
be included in the relationship, i.e., $\log (T_{\rm B}^{L_{\rm X}}/
{\rm day)}=2.1\log (\bh/10^6\msun) - 0.98 \log (L_{\rm bol}/10^{44}
{\rm erg~s}^{-1}) - 2.3$, where $L_{\rm bol}$ is the bolometric
luminosity; Gonz{\'a}lez-Mart{\'{\i}}n \& Vaughan (2012) recently obtained a similar
relationship, i.e., $\log (T_{\rm B}^{L_{\rm X}}/ {\rm day)}=1.34\log
(\bh/10^6\msun) - 0.24 \log (L_{\rm bol}/10^{44}~{\rm
erg~s}^{-1})-1.9$, of which the dependence on the accretion rate is
subsequently weaker compared with that in \citet{McHardy06}. 

To study the scaling relation between $T_{\rm B}^{L_{\rm X}}$ and
$\bh$ in the absorption scenario for the X-ray variations, we perform
more simulations for those models with the same settings as 
models A1, C1, and C2, respectively, except for adopting various masses
for the central MBH in the range from  $10^6$ to $2\times10^8\msun$.
(For other models, we obtain the similar conclusions.) We find that  the shape
of the PSD is independent of the setting for the hydrogen densities of those
clouds with semimajor axis of $a_{\rm c,min}$ (i.e., $n_{\rm
H,\bh,0}$) if the absorption column density is in the range from $\sim
10^{21} $ to $10^{24}~\cm^{-2}$. Therefore, we simply set $n_{\rm
H,\bh,0}=n_{\rm H,0}(10^7\msun/\bh)$, where $n_{\rm H,0}$ is the initial
setting for those cases with $\bh=10^7\msun$ as listed in
Table~\ref{tab:t1}. With this setting, those cases with too many Compton-thick
clouds are avoided.
Alternatively setting an $n_{\rm H,0}$ a few
times larger, or smaller, for those models listed in Table 1, does not
significantly affected the model results.

Figure~\ref{fig:f10} shows our model results on $T_{\rm B}^{L_{\rm
X}}$ as a function of $\bh$. As seen from the figure, the absorption scenario for
the AGN X-ray variations naturally leads to a strong correlation
between $T_{\rm B}^{L_{\rm X}}$ and $\bh$, which can be consistent
with the relationship obtained from observations. The scatters in the
spatial distribution and the intrinsic properties of the eclipsing
clouds among individual AGNs, indicated by the different model
settings listed in Table~\ref{tab:t1}, and those shown in
Figure~\ref{fig:f9}, may lead to a scatter in the relationship as
shown in Figure~\ref{fig:f10}.  

The rough consistency between the observations and our model results
suggests that the absorption model, i.e., that the AGN X-ray variations are
mainly caused by those eclipsing clouds crossing the LOS, may provide
a natural explanation to the shapes of the PSDs for some, if not all,
AGNs and the scaling relation between the break timescales and the
MBH masses for those AGNs.  In principle, a comparison between the
$T^{L_{\rm X}}_{\rm B}-\bh$ relation resulting from those absorption
models and the observational ones may put strong constraints on the
model parameters. However, it appears there are large uncertainties in
the observationally estimated $T^{L_{\rm X}}_{\rm B}-\bh$ relation (as
large as $1$\,dex), which prevent a detailed statistical comparison of
the model results with the observational ones.

For those absorption models with the same settings on the spatial
distribution and the intrinsic properties of the clouds but with different
masses of the central MBH, we find that the slopes of the resulting
PSDs at both the low frequencies ($\gamma_{\rm l}$) and the high frequencies
($\gamma_{\rm h}$) are independent of the MBH mass (see the right
panel of Figure~\ref{fig:f10}). No strong dependence of the scaling
relation on the Eddington ratio is found for the models studied in
this paper because only one model parameter, $r_{\rm t}$, is
set to correlate with the Eddington ratio. 

Note here that in the absorption scenario a correlation between
$T^{L_{\rm X}}_{\rm B}$ and $\bh$ with a slope of about 1 or
$1+\frac{1}{2}(\alpha_{R_{\rm c}}^{-1}-1)$
in the log-log space can be produced if $R_{\rm X}$ and
$a_{\rm c,min}$ scale linearly with the MBH mass $\bh$. It is
plausible that $R_{\rm X}$ and $a_{\rm c,min}$ scale linearly with
the MBH mass $\bh$, since the intrinsic properties of the clouds and
the spatial distribution of those clouds probably scale with the
luminosity of an AGN, which is proportional to the Eddington
luminosity and thus linearly scales with the MBH mass. 
In reality, it is possible that other parameters, e.g., the cloud radius
$R_{\rm c}$ and
the inner boundary for the spatial distribution of eclipsing clouds
$a_{\rm c,min}$, are correlated with the Eddington ratio
and consequently the break timescale (or frequency) also depends on
the Eddington ratio. Since the relationships between these parameters
and the Eddington ratio are not clear, a further exploration of
the dependence of the scaling relation on the Eddington ratios is beyond the
scope of the paper.

\subsection{On the Amplitude of the Variation}
\label{subsec:var}

The variation of the X-ray emission from an AGN can be quantified by
the normalized excess variance (NEV) 
\be 
\sigma_{{\rm NEV},L}^2=\frac{\displaystyle{\sum^N_{i=1}} (L_{{\rm
X}}(t_i)-\langle L_{\rm X}\rangle)^2}{N\langle L_{\rm X}\rangle^2}.
\label{eq:spc_var} 
\ee
If the variation of
the X-ray emission of some AGNs is mainly due to the absorption by
eclipsing clouds as assumed in this paper, the NEV ($\sigma_{{\rm
NEV},L}$) for those models listed in Table~\ref{tab:t1} can be
obtained from our simulations. Similarly, the NEV of the absorption
column density $\sigma_{{\rm NEV},N_{\rm H}}$ can be obtained by
replacing $L_{\rm X}(t_i)$ and $\langle L_{\rm X}\rangle$ by $N_{\rm
H}(t_i)$ and $\langle N_{\rm H}\rangle$ in Equation (\ref{eq:spc_var}),
respectively, and $N_{\rm H}(t_i)$ are obtained from the mock
observations at the $i$-th time interval. The actual value of
$\sigma_{\text{NEV},L}^2$ depends on the total observational time
$T_{\text{tot}}$ and the duration of the observational time interval
$\delta t$. If $T_{\text{tot}}$ is not sufficiently large, then those
eclipses caused by clouds with large $a_{\text{c}}$ cannot be fully
counted, and if $\delta t$ is too large, then those eclipsing clouds
with small $a_{\text{c}}$ are also not fully counted.

We assume that the AGN X-ray variations for different AGNs follow an intrinsic
PSD with the following universal double power-law form, i.e., 
\be
P(\nu) \simeq
\begin{cases} 
A \left(\frac{\nu}{\nu_{\rm B}}\right)^{\gamma_{\rm h}}, \quad
\text{when } \nu \geq \nu_{\rm B}, \\ 
A \left(\frac{\nu}{\nu_{\rm B}}\right)^{\gamma_{\rm l}}, \quad
\text{when } \nu < \nu_{\rm B}  \\
\end{cases}
\label{eq:pnu2}
\ee 
as a simplified form of Equation~(\ref{eq:bkpow}), where $A$ is the
amplitude of the PSD at $\nu_{\rm B}$, $\gamma_{\text{h}} <
\gamma_{\text{l}}$, and $\gamma_{\text{h}} <-1$. If the frequency
range limited by ``observations'' is from $\nu_{\rm min} \sim 1/T_{\rm
tot}$ to $ \nu_{\text{max}} \sim 1/(2\delta t)$ (which is substantially
narrower than the range considered in the models), and
$\nu_{\text{min}} \ll \nu_{\text{max}}$, then the NEV estimated from
the observations can be approximately given by 
\begin{eqnarray}
& &\sigma^2_{\text{NEV},L} \simeq
\int^{\nu_{\text{max}}}_{\nu_{\text{min}}} 
\delta_L^2
P(\nu) d\nu =
\delta_L^2
\frac{A\nu_{\text{B}}}{\gamma_{\text{h}}+1} \times  \nonumber \\
& & \begin{cases}
\left[
\left(\frac{\nu_{\text{max}}}{\nu_{\text{B}}}\right)^{\gamma_{\text{h}}+1} 
- \left(\frac{\nu_{\text{min}}}{\nu_{\text{B}}}\right)^{\gamma_{\text{h}}+1}
  \right], \\ ~~~~~~~~~~~~~~~~~~~~~~~~~~~~~~~~~~~~~~~~\text{when }
\nu_{\text{min}} \geq \nu_{\text{B}}; \\
\left[
\left(\frac{\nu_{\text{max}}}{\nu_{\text{B}}}\right)^{\gamma_{\text{h}}+1}-1\right]+
\frac{\gamma_{\text{h}}+1}{\gamma_{\text{l}}+1}
\left[1-\left(\frac{\nu_{\text{min}}}{\nu_{\text{B}}}\right)^{\gamma_{\text{l}}+1}
\right], \\ ~~~~~~~~~~~~~~~~~~~~~~~~~~~~~~~~~~~~~~~~\text{when }
\nu_{\text{min}} <\nu_{\text{B}} < \nu_{\text{max}}; \\
\frac{\gamma_{\text{h}}+1}{\gamma_{\text{l}}+1} \left[
\left(\frac{\nu_{\text{max}}}{\nu_{\text{B}}}\right)^{\gamma_{\text{l}}+1} 
- \left(\frac{\nu_{\text{min}}}{\nu_{\text{B}}}\right)^{\gamma_{\text{l}}+1}
  \right], \\ ~~~~~~~~~~~~~~~~~~~~~~~~~~~~~~~~~~~~~~~~\text{when }
\nu_{\text{max}} \leq \nu_{\text{B}},  ~\label{eq:sigma_nevl}
\end{cases}
\end{eqnarray}
where
$\delta_L^2\equiv\mathop{\lim}\limits_{\Delta T\rightarrow\infty} \int_0^{\Delta T}
[1-L(t)/\overline{L(t)}]^2 dt/\Delta T$ and $\overline{L(t)}\equiv\mathop{\lim}\limits_{\Delta T\rightarrow\infty} 
\int_0^{\Delta T} L(t)dt/\Delta T$.
If $\gamma_{\text{h}}$ [or $\gamma_{\text{l}}$] equals
$-1$, the terms
$\left(\nu_{\text{min}}/\nu_{\text{B}}\right)^{\gamma_{\text{h}}+1}$
and
$\left(\nu_{\text{max}}/\nu_{\text{B}}\right)^{\gamma_{\text{h}}+1}$
[or
$\left(\nu_{\text{min}}/\nu_{\text{B}}\right)^{\gamma_{\text{l}}+1}$
and
$\left(\nu_{\text{max}}/\nu_{\text{B}}\right)^{\gamma_{\text{l}}+1}$]
should be replaced by $(\gamma_{\text{h}}+1)\ln
\left(\nu_{\text{min}}/\nu_{\text{B}}\right)$ and
$(\gamma_{\text{h}}+1)\ln
\left(\nu_{\text{max}}/\nu_{\text{B}}\right)$ [or
$(\gamma_{\text{l}}+1)\ln
\left(\nu_{\text{min}}/\nu_{\text{B}}\right)$ and
$(\gamma_{\text{l}}+1)\ln
\left(\nu_{\text{max}}/\nu_{\text{B}}\right)$], respectively.

In the absorption scenario presented in this paper, the X-ray
variations are due to the eclipses of clouds in the BLR and the dusty
torus. As shown in Figure~\ref{fig:f10}, the power at
$\nu_{\text{B}}$, i.e., $A\nu_{\text{B}}$, is more or less a constant
for the different BH masses of a given model and $\nu_{\text{B}} \propto \bh^{-1}$. If
$\delta_L^2$ is also constant with different $\bh$, then $\sigma^2_{\text{NEV},L}
\propto \bh^{\gamma_{\text{h}}+1}$ when
$
\nu_{\text{min}}
>\nu_{\text{B}}
$,
and $\sigma^2_{\text{NEV},L}\propto \bh^{\gamma_{\text{l}}+1}$ when
$\nu_{\text{max}} <\nu_{\text{B}}$, and transits from $\propto
\bh^{\gamma_{\text{h}}+1}$ to $\propto \bh^{\gamma_{\text{l}}+1}$
when $\nu_{\text{B}}$ moves from $\nu_{\text{min}}$ to $\nu_{\text{max}}$. This suggests that the absorption scenario for the
X-ray variations can also naturally lead to a correlation between the
MBH mass and the NEV.  For example, model C2 results in a PSD with
$\gamma_{\text{h}}=-2.6$ and $\gamma_{\text{l}}=-0.9$, which may lead
to a correlation $\sigma^2_{\text{NEV},L} \propto \bh^{-1.6}$ at the
high-MBH mass range and $\propto \bh^{0.1}$ at the low-MBH mass range.

A scatter in the $\sigma^2_{\text{NEV}}$-$\bh$ correlation can be caused
at the least by the following factors. 
(1) In reality, $\delta_L^2$ may depend on the settings of the
spatial distribution and the intrinsic properties of the eclipsing clouds,
which could lead to a scatter in the $\sigma^2_{\text{NEV}}$-$\bh$ correlation.
For example, we find that $\delta_L^2 \propto \left< N \right>_{\text{LOS}}$;
even if the cloud size and the size of the X-ray emitting region scale
linearly with the MBH mass to cancel out some dependence on the MBH mass
in Eq.~(\ref{eq:nlos}), it still depends on the total
number of clouds $N_{\text{tot}}$ and the viewing angle.
(2) The differences among the spatial distributions of the eclipsing clouds in
different AGNs with the same MBH mass may result in different shapes of the
PSDs (defined by $\nu_{\text{B}}$, $\gamma_{\text{h}}$ and
$\gamma_{\text{l}}$), which consequently leads to a scatter to the
$\nu_{\text{B}}$-$\bh$ relation and thus further introduce scatters to the
$\sigma^2_{\text{NEV}}$-$\bh$ relation.

Note that a number of studies have shown that the magnitudes of the AGN
X-ray variation on short timescales tightly correlate with the masses
of the central MBHs \citep[e.g.,][]{LY01b, Oneil05, Nikoajuk09, Zhou10,
Ponti12, McHardy13, Soldi14}. The variation magnitudes on long
timescales may saturate and become independent of the MBH  masses
\citep{Shimizu13, Markowitz04}.  These observations can be well
interpreted as the X-ray variations of different AGNs following a
universal PSD, with a break frequency correlating with the MBH mass
\citep{Markowitz04}.  This relation may be a direct result of a
uniform PSD with $\gamma_{\text{h}} \sim -4$---$-2$ and $\gamma_{\rm l}$ close
to -1 for the X-ray variation
in those AGNs as discussed intensively in the literature
\citep[e.g.,][]{Ponti12,Ludlam15,Pan15}.

\section{Conclusions and Discussions}
\label{sec:discussion}

In this paper, we study the AGN X-ray variations due to the absorption of
the clouds or clumps in the BLR and the dusty torus that happen to
cross the LOS. In this absorption scenario for the AGN X-ray
variations, we investigate the dependence of the power spectral
densities (PSDs) of the X-ray flux and the absorption column density
variations on the spatial distribution and the intrinsic properties of
those clouds. We analyze various statistical properties of the X-ray
eclipsing events, e.g., the event rate, the mean number of the
eclipsing clouds per unit time, and the statistical distributions of
the parameters for the eclipsing events. We perform Monte-Carlo
simulations to realize the kinematics of those clouds in the vicinity
of AGNs and obtain mock X-ray variations due to the X-ray eclipses. We
find that the resulting PSDs of the X-ray flux or the absorption
column density variations can be described by a breaking double
power-law form in the frequency range from $10^{-3}$\,Hz to
$10^{-9}$\,Hz, which can be well consistent with those measured from
observations.  The PSD at the low (or high) frequencies is mainly
controlled by the spatial distribution and the intrinsic properties of
the eclipsing clouds at the outer (or inner) region, presumably by some
clouds in the dusty torus (or BLR). 

We find that the break frequency of the PSD is roughly determined by
either the eclipsing durations of those eclipsing clouds with the
minimum semimajor axis (i.e., $\nu \sim 1/t_{\rm E'}(a_{\rm c,min})$)
or the eclipsing duration of those clouds with a characteristic
semimajor axis ($a_{\rm eq}$, which is defined by that the sizes of
the clouds with semimajor axes larger than $a_{\rm eq}$ are larger
than the size of the X-ray emission region, while the sizes of the
clouds with semimajor axes smaller  than $a_{\rm eq}$  are smaller
than the size of the X-ray emission region). We demonstrate that the
break timescales, corresponding to the break frequencies of the
PSDs, are strongly correlated with the masses of the central MBHs in
the cloud absorption scenario for the X-ray variations of AGNs, which
may provide a natural explanation to the scaling relation suggested by
observations for some AGNs \citep{Markowitz03, McHardy06,
GonzlezMart12}.  If future observations can more accurately determine
this scaling relation, the cloud absorption scenario for the X-ray
variations of AGNs may be further constrained and the underlying
physics for the scaling relation may be better understood. This
scaling relation is, therefore, expected to provide a robust tool to
estimate the masses of MBHs in some type 2, if not all, AGNs. (Note
that the popular reverberation mapping technique is not easy to 
apply to type 2 AGNs, for which no strong broad emission lines can
be directly detected.) For some type 1 AGNs, this relation  may also be
applicable to estimate the masses of the central MBHs if their
X-ray variations are dominated by the absorption of eclipsing clouds.
We also show that this correlation, together with the assumption that the
X-ray variation of different AGNs follows a universal PSD, will lead to
a dependence of the X-ray variation amplitude on BH mass as shown in some
observations.

The spatial distribution and the intrinsic properties of the eclipsing
clouds and their parent population can be extracted from the X-ray
flux variation (if it is mainly due to the absorption of eclipsing
clouds), the column density variation, and the PSDs of these
variations. Observations have shown that the X-ray variations of some
AGNs, such as NGC 1365~\citep[e.g.][]{Risaliti99}, NGC
7582~\citep[e.g.][]{Risaliti02} , and NGC 4151~\citep{Schurch02,
Markowitz14}, are probably dominated by the absorption of eclipsing
clouds.  For those AGNs, it is possible to adopt the X-ray eclipse
model introduced in this study to match the PSDs of their X-ray flux
variations (or the PSDs of their absorption column density variations,
if available) individually.  With such a modeling, robust constraints
may be obtained on the spatial distribution and the intrinsic
properties of the eclipsing clouds and their parent populations for
those individual AGNs. The statistical scatters of the spatial
distribution and the intrinsic properties of the clouds in different
AGNs and its dependence on various AGN properties, such as luminosity,
MBH mass, Eddington ratio, etc., may also be revealed by studying a
sample of such AGNs. This may help to establish a unified theory for
the nature of the BLR and the torus. 

As the UV-optical sources (mainly from the accretion disk) are
much more extended than the X-ray sources, it is likely that some AGNs show
X-ray variations due to eclipsing clouds, while they appear as type 1 AGNs in
the UV-optical band. For these objects, it will be interesting to investigate
the flux variability in the UV-optical band due to the covering of the clouds
together with the variation of the X-ray emission. The comparison of both the
model predictions and observations in the multiple bands can help to set robust
constraints on the structure and property distributions of the AGN clouds, and
also the type 1 and type 2 dichotomy. Note that the covering fraction of the
clouds and the variability of the flux appearing in the UV-optical band may
have a different dependence on the cloud structure and properties from those in
the X-ray band.  A comprehensive exploration of these features, which is beyond
the scope of this paper, is worthy. 

In addition to X-ray flux variations, the
variation of the Fe K$\alpha$ line emission due to eclipsing may also be
used to probe the kinematic structure of the parent population of
the eclipsing clouds and the inner disk structure.

It has been shown that the spatial distribution and the intrinsic
properties of the BLR clouds and the clumps in the dusty torus can be
revealed by the reverberation mapping technique
\citep[e.g.,][]{Li13,Chelouche13,Pancoast11,Pancoast13}. If the X-ray
variations in some AGNs with broad emission lines are due to the
absorption of eclipsing clouds in the BLR and/or the dusty torus, it
may be possible to combine both the reverberation mapping technique
and the PSD analysis of the X-ray flux (and the absorption column
density) variations to study and constrain the properties of those
clouds in the BLR and the dusty torus, which may help to reach a
coherent understanding of various AGN emission/absorption features and
the immediate environment of the central engine of AGNs.

For some AGNs, the observed X-ray flux variations may be mainly due to
the changes of the properties of the inner accretion disk and the
X-ray emitting corona  \citep[e.g.,][]{Lyubarskii97, Lamer03a,
Uttley02, Zdziarski03, PF99, LY01a, Fabian03, Marinucci14}, of which
the PSDs of the X-ray variations may be quite different from those
obtained from the absorption scenario studied in this paper. If it is
due to the flicker noise in accretion disks as suggested by
\citet{Lyubarskii97}, for example, the resulting PSD may be typically
$\propto \nu^{-1}$. In the literature, however, there are no explicit
predictions on the PSD of the X-ray flux variations due to the
intrinsic changes for most of other proposed models. Future progress on
the studies of the intrinsic variation of the X-ray emission from AGNs
may provide some information on the PSD of this variation, which may
be used to distinguish from the absorption scenario for the X-ray flux
variation of some AGNs.

Note here that we mainly focus on the variations of the X-ray emission
in the $2-10$\,keV band in this paper. For the X-ray emission at a
band with higher energy, e.g., $10-100$\,keV, the shape of the
resulting PSD is the same as that for the $2-10$\,keV X-ray emission,
although the amplitude of the variation at the high-energy band is
substantially smaller than that at the $2-10$\,keV band. The
independence of the PSD shape from the energy and the dependence of the
variation amplitude on the energy are the simple nature of the
absorption scenario for the X-ray variations, which may be used to
distinguish the absorption scenario from those models assuming
intrinsic variations \citep[e.g.,][]{Miller08}.

\acknowledgements
\noindent This work was supported in part by the National Natural Science
Foundation of China under nos.\ 11373031, 11390372, 11273004, 11603083, the
National Key Program for Science and Technology Research and Development
(grant Nos.  2016YFA0400703, 2016YFA0400704), the Strategic Priority
Program of the Chinese Academy of Sciences (grant No. XDB 23040100), and the
Fundamental Research Funds for the Central Universities grant No. 161GPY51.
F.Z. is partly supported by a postdoctoral fund 2014M550549.

\end{document}